\documentclass[journal]{IEEEtran}

\ifCLASSINFOpdf

\else

\fi

\usepackage{mathrsfs}
\usepackage{amsmath}
\usepackage{amsfonts}
\usepackage{amsthm}
\usepackage{amssymb}
\usepackage{graphicx}
\usepackage{subfigure}
\usepackage{indentfirst}
\usepackage{array}
\usepackage{epstopdf}
\usepackage{cite}
\usepackage{mathrsfs}
\usepackage{enumerate}
\usepackage{bm}
\usepackage[linesnumbered,ruled,vlined]{algorithm2e}
\usepackage{setspace}
\usepackage{multirow}
\usepackage{verbatim}
\usepackage{stfloats}
\usepackage{color}
\usepackage[table]{xcolor}
\usepackage{makecell}
\usepackage{cancel}
\usepackage{threeparttable}
\usepackage{pifont}
\SetKwComment{Comment}{//}{}

\begin{document}

\title{Toward Adaptive Semantic Communications: Efficient Data Transmission via Online Learned Nonlinear Transform Source-Channel Coding}

\author{Jincheng~Dai,~\IEEEmembership{Member,~IEEE},
        Sixian~Wang,~\IEEEmembership{Graduate Student Member,~IEEE},
        Ke~Yang,~\IEEEmembership{Graduate Student Member,~IEEE},
        Kailin~Tan,~\IEEEmembership{Graduate Student Member,~IEEE},
        Xiaoqi~Qin,~\IEEEmembership{Member,~IEEE},
        Zhongwei~Si,~\IEEEmembership{Member,~IEEE},
        Kai~Niu,~\IEEEmembership{Member,~IEEE},
        and Ping~Zhang,~\IEEEmembership{Fellow,~IEEE}

\thanks{This work was supported in part by the National Natural Science Foundation of China under Grant 62293481, Grant 92067202, Grant 62001049, Grant 62071058, and Grant 61971062, in part by the Beijing Natural Science Foundation under Grant 4222012, in part by Program for Youth Innovative Research Team of BUPT No. 2023QNTD02. \emph{(Corresponding authors: Jincheng Dai, Ping Zhang)}}

\thanks{Jincheng Dai, Sixian Wang, Ke Yang, Kailin Tan, and Zhongwei Si are with the Key Laboratory of Universal Wireless Communications, Ministry of Education, Beijing University of Posts and Telecommunications, Beijing 100876, China (e-mail: daijincheng@bupt.edu.cn).}

\thanks{Xiaoqi Qin, Kai Niu, and Ping Zhang are with the State Key Laboratory of Networking and Switching Technology, Beijing University of Posts and Telecommunications, Beijing 100876, China.}

\vspace{-1em}
}

\maketitle

\begin{abstract}
	The emerging field \emph{semantic communication} is driving the research of end-to-end data transmission. By utilizing the powerful representation ability of deep learning models, learned data transmission schemes have exhibited superior performance than the established source and channel coding methods. While, so far, research efforts mainly concentrated on architecture and model improvements toward a static target domain. Despite their successes, such learned models are still suboptimal due to the limitations in model capacity and imperfect optimization and generalization, particularly when the testing data distribution or channel response is different from that adopted for model training, as is likely to be the case in real-world. To tackle this, in this paper, we propose a novel online learned joint source and channel coding approach that leverages the deep learning model's overfitting property. Specifically, we update the off-the-shelf pre-trained models after deployment in a lightweight online fashion to adapt to the distribution shifts in source data and environment domain. We take the overfitting concept to the extreme, proposing a series of implementation-friendly methods to adapt the codec model or representations to an individual data or channel state instance, which can further lead to substantial gains in terms of the end-to-end rate-distortion performance. Accordingly, the streaming ingredients include both the semantic representations of source data and the online updated decoder model parameters. The system design is formulated as a joint optimization problem whose goal is to minimize the loss function, a tripartite trade-off among the data stream bandwidth cost, model stream bandwidth cost, and end-to-end distortion. The proposed methods enable the communication-efficient adaptation for all parameters in the network without sacrificing decoding speed. Extensive experiments, including user study, on continually changing target source data and wireless channel environments, demonstrate the effectiveness and efficiency of our approach, on which we outperform existing state-of-the-art engineered transmission scheme (VVC combined with 5G LDPC coded transmission).
\end{abstract}


\IEEEpeerreviewmaketitle

\section{Introduction}\label{section_introduction}

\IEEEPARstart{S}{emantic} communications are recently emerging as a new paradigm driving the in-depth fusion of information and communication technology (ICT) advances and artificial intelligence (AI) innovations \cite{guduz2023beyond, zhang2021toward,xie2021deep,dai2021semantic}. Unlike traditional communication design philosophy that focuses on accurately transmitting bits over a noisy communication channel \cite{Shannon1948A}, semantic communications are goal-oriented, which helps the transceiver identify the most valuable information more efficiently, i.e., the information necessary to recover the purpose intended by the transmitter. Performance assessment also goes beyond the common Shannon paradigm of guaranteeing the correct reception of each single transmitted bit, human perceptual loss \cite{zhang2018unreasonable,ding2020image,heusel2017gans} and machine task accuracy \cite{duan2020video} are taken as the distortion, which is better aligned with the essential goal of end-to-end communications \cite{guduz2023beyond}.

One roadmap to realize semantic communication is bridging the source and channel parts together to boost the end-to-end content delivery. The paradigm aiming at the integrated design of source and channel processing is \emph{joint source-channel coding (JSCC)} \cite{verduJSCC}, a classical topic in the information theory and coding theory. However, conventional JSCC schemes \cite{verduJSCC,guyader2001joint,ramzan2007joint,chen2018joint} are limited by explicit probabilistic models and handcrafted designs, whose optimization is intractable for complex sources. They also ignore the semantic feature aspects of source messages and cannot be optimized towards human perception or machine task directly. In contrast, the emerging semantic communication utilizes deep learning models to realize JSCC \cite{DJSCC,DJSCCF,DJSCCL,jankowski2020wireless,JSCCtext,choi2019neural,xu2021wireless,tung2021deepwive}, which can be optimized for specific end-to-end transmission objectives. For example, in the case of wireless image transmission, deep JSCC approaches have been verified to surpass classical separation-based BPG source compression \cite{BPG} combined with advanced low-density parity-check (LDPC) channel coding \cite{richardson2018design}, especially for sources of tiny dimensions, e.g., CIFAR dataset ($32 \times 32$ pixels). To address the challenge of high resolution media transmission, the optimization goal of JSCC system should be formulated as a trade-off between the reconstruction quality (end-to-end distortion) and the channel bandwidth cost (channel bandwidth ratio). Following this, nonlinear transform source-channel coding (NTSCC) proposed in \cite{dai2022nonlinear} has achieved content-aware variable-length JSCC via introducing an entropy model on the semantic latent representations, which can significantly improve the overall coding efficiency. This coding paradigm reveals the key aspects of semantic transmission: maximizing reconstruction quality meets the human perception; whereas minimizing wireless channel bandwidth cost benefits the efficient transmission, i.e., \emph{channel bandwidth ratio-distortion (CBR-D) trade-off optimization}.

Although existing end-to-end transmission approaches have proven to be successful in optimizing the end-to-end \emph{expected CBR-D} trade-off over a source dataset and ergodic wireless channel responses, they are yet unlikely to be optimal for every test instance due to the limited model capacity and imperfect optimization. In essence, they assume that if a model performs well on both training and validation datasets, it will likely generalize well to new, unseen data and channel responses during testing. However, this assumption does not always hold in practice. On the one hand, such an amortized model might not be good at capturing the data semantic feature and channel state for each instance, resulting in suboptimal transform and coding during the inference stage. On the other hand, the imperfect optimization and generalization will be especially severe when the testing data distribution or channel response is different from that adopted in the training stage. To tackle this, we explore a new online learned approach by optimizing the network parameters or the semantic representations during the model inference stage, based on the current target source data and wireless channel domain. In other words, we turn to optimize the CBR-D trade-off for substantial gains on every data and channel state instance.

Some insights in traditional source compression codec can bring some inspirations \cite{gu2011low,lan2017variable,puri2016annealed}. Conventional image/video compressors follow the hybrid transform coding paradigm. For example, HEVC \cite{sullivan2012overview} and VVC \cite{bross2021developments} jointly use discrete cosine transform (DCT) and discrete sine transform (DST) to handle different types of signals. Multiple transform selection (MTS) mechanism is introduced to VVC standard to select the most appropriate transform locally aiming at the best rate-distortion trade-off. Inspired by the idea of signal-dependent transform in traditional source compression methods, neural codecs are also evolving toward the \emph{instance-adaptive} paradigm. The suboptimality of neural codecs has been studied extensively in terms of the model inference suboptimality \cite{cremer2018inference}. It has been shown that by online finetuning the encoder parameters or latent features from a well-trained model for a particular instance, substantial gain can be further obtained in the compression rate-distortion performance \cite{lu2020content,campos2019content,yang2020improving,guo2020variable,kim2020neural,liu2021overfitting}. Some new methods are recently developed to further improve the rate-distortion cost, e.g., a full-model instance-adaptive compression method was proposed in \cite{van2021overfitting}, a neural syntax method was proposed in \cite{wang2022neural} to realize data-dependent compression, etc. All these methods leverage neural network's overfitting property, adapting the model to an individual source data sample.

\begin{figure}[t]
	\centering{\includegraphics[scale=0.24]{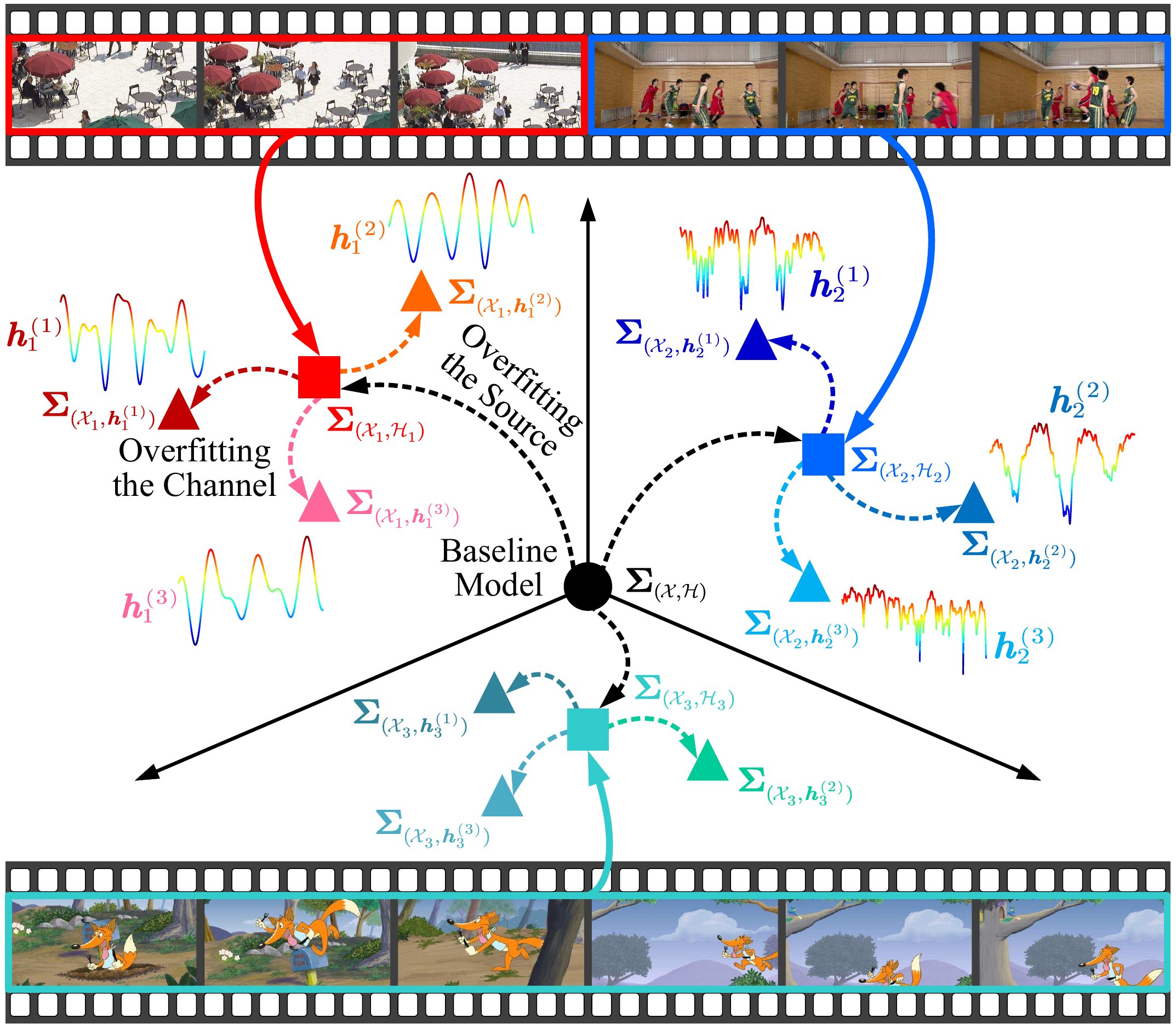}}
	\caption{We first adapt the baseline model $\boldsymbol{\Sigma}_{(\mathcal{X},\mathcal{H})}$ learned from the source data set $\mathcal{X}$ and CSI set $\mathcal{H}$ to the given specific source domain $\mathcal{X}_k$ and CSI domain $\mathcal{H}_k$, i.e., overfitted models $\boldsymbol{\Sigma}_{(\mathcal{X}_k,\mathcal{H}_k)}$. To address the problem of transmission over the time-varying wireless channel, we further design a channel-dependent method to adapt the model $\boldsymbol{\Sigma}_{(\mathcal{X}_k,\mathcal{H}_k)}$ to every specific CSI instance $\boldsymbol{h}_k^{(i)} \in {\mathcal{H}_k}$, resulting in the final overfitted model $\boldsymbol{\Sigma}_{(\mathcal{X}_k,{\boldsymbol{h}}_k^{(i)})}$.}\label{Fig_paper_idea}
	\vspace{-1em}
\end{figure}

Inspired by the useful insights from traditional compression codecs and neural compression codecs, in this paper, we make the first attempt to build a neural instance/domain adaptive joint source-channel coding architecture for end-to-end data transmission. We show that overfitted neural-enhancement on end-to-end communication systems is feasible and effective when combined with online learning. Accordingly, we propose \emph{online learned adaptive NTSCC}, a novel framework that employs online learning to overfit the instant source data sample and channel state information (CSI). Our new method is easy to implement and can be incorporated in a number of different architectures of semantic communications. In this paper, as a representative case, we inject our online adaptation method to the nonlinear transform source-channel coding (NTSCC) based semantic communication system \cite{dai2022nonlinear}, to validate the effectiveness and efficiency. In contrast to previous works, our system introduces an additional \emph{model stream} alongside the traditional \emph{data stream}, which is utilized to update the JSCC decoder and synthesis transform parameters at the receiver. We take into account the costs of sending the model updates such that the whole system design is formulated as an optimization problem whose goal is to minimize the loss function that is a tripartite trade-off among the data stream bandwidth cost ($R$), model stream bandwidth cost ($M$), and end-to-end distortion ($D$) terms. Our proposed method is lightweight, effective and intuitively appealing, which could be feasibly transplanted to any existing deep learning based semantic communication systems. Fig. \ref{Fig_paper_idea} illustrates the key idea of our work briefly.

Specifically, the contributions of this paper can be summarized as follows.

\begin{enumerate}[(1)]
	\item \emph{Online Learned NTSCC Framework:} We make the first attempt to build a source and channel instance or domain adaptive semantic communication system based on the online learned NTSCC. Our innovative overfitting mechanism enables the whole NTSCC system to be much more powerful and flexible for extracting more compact semantic representations, offering superior end-to-end bandwidth ratio-distortion performance. We make variational analysis to interpret the origins of general model inference suboptimality, and clarify the rationale of our proposed source and channel overfitting paradigm.
	
	\item \emph{Overfitting the Source:} We design two simple yet efficient ways to realize the source data instance or domain adaptive NTSCC. These methods are of different ideas for adapting a given pre-trained model or semantic latent representation to an individual data sample or other content domain that is different in appearance. We discuss the specific scenario for which each method is suitable. Our source overfitting mechanism online adapts a baseline model to each specific scene to maximize communication efficiency.
	
	\item \emph{Overfitting the Channel:} We propose a plug-in CSI modulation module inserted into pre-trained codec modules. It enables the whole system to efficiently deal with different channel states with a single trained network. The proposed scheme can not only adapt to different signal-to-noise ratio (SNR) under the block fading channel, but also provide consistent and robust performance under diverse frequency selective fading channels, which makes our model agilely transferred over various channel states.
	
	\item \emph{Performance Validation:} We verify the effectiveness and efficiency of online learned NTSCC over video I-frame sources and practical wireless channels. Extensive results indicate that our method can lead to substantial gains in the CBR-D performance without sacrificing decoding speed. Equivalently, achieving the same end-to-end transmission performance, the proposed transceiver adaptation scheme can save up to 45\% bandwidth cost compared to the state-of-the-art (SOTA) engineered transmission scheme (VVC combined with 5G LDPC coded transmission).
\end{enumerate}

The remainder of this paper is organized as follows. In Section \ref{section_preliminaries}, we review the architecture and properties of NTSCC system, and analyze its suboptimality using the variational inference. Next, in Section \ref{section_source_overfit}, we present our source overfitting methods, including different ideas for online overfitting a given pre-trained baseline model to an individual data sample or other content domain. In Section \ref{section_channel_overfit}, we show our channel overfitting methods, including details on a plugin-in channel modulation module to adapt a pre-trained model to the instant channel state. Section \ref{section_performance} shows experimental results to quantify our performance gain, and some valuable discussions are also given. Finally, Section \ref{section_conclusion} concludes this paper.

\emph{Notational Conventions:} Throughout this paper, lowercase letters (e.g., $x$) denote scalars, bold lowercase letters (e.g., $\boldsymbol{x}$) denote vectors. In some cases, $x_i$ denotes the elements of $\boldsymbol{x}$, which may also represent a subvector of $\boldsymbol{x}$ as described in the context. Bold uppercase letters (e.g., $\boldsymbol{X}$) denote matrices, and $\boldsymbol{I}_m$ denotes an $m$-dimensional identity matrix. $\ln (\cdot)$ denotes the natural logarithm, and $\log (\cdot)$ denotes the logarithm to base $2$. $p_x$ denotes a probability density function (pdf) with respect to the random variable $x$. In addition, $\mathbb{E} (\cdot)$ denotes the statistical expectation operation, and $\mathbb{R}$ denotes the real number set. Finally, $\mathcal{N}(x|\mu, \sigma^2) \triangleq (2\pi \sigma^2)^{-1/2} \exp(-(x - \mu)^2/(2\sigma^2))$ denotes a Gaussian function, and $\mathcal{U}(a-u,a+u)$ stands for a uniform distribution centered on $a$ with the range from $a-u$ to $a+u$.

\section{Preliminaries and Motivation}\label{section_preliminaries}

Built upon the variational auto-encoder (VAE) \cite{kingma2013auto} architecture, NTSCC has shown superior performance on wireless image and video transmission problems \cite{guduz2023beyond,dai2022nonlinear,wang2022wireless}. Owing to the powerful ability of representation learning, NTSCC can well extract the source semantic features and transmit them over the wireless channels efficiently by using variable-length deep JSCC techniques. This method not only achieves comparable or better performance than the SOTA engineered source compression combined with advanced channel coding schemes, but also greatly surpasses plain auto-encoder based deep JSCC methods \cite{DJSCC} that directly encode the raw source data rather than its semantic features. In addition, NTSCC shows great potential to achieve lower time complexity due to its efficient parallel computing with deep neural networks (DNNs). The above superiority of NTSCC can well support semantic communications. Therefore, in this paper, we choose NTSCC to build the semantic communication system. Similar to previous works, we take image or video I-frame (intra-coded frame) source as representative, but our work is extensible for other source modalities.

\subsection{NTSCC based Semantic Communication System}

The idea of NTSCC stems from the landmark work of Ball\'{e} \emph{et al.} on nonlinear transform coding (NTC) \cite{balle2020nonlinear,balle2016,balle2018}. Given the pristine data sample $\boldsymbol{x}$, e.g., an image $\boldsymbol{x}$ modeled as a vector of pixel intensities $\boldsymbol x \in {\mathbb R}^m$, it is first transformed into semantic latent representation $\boldsymbol{y}$ using a DNN-based nonlinear analysis transform $g_a$. In data compression tasks, $\boldsymbol{y}$ will be quantized as discrete-valued latent representation $\boldsymbol{\bar y}$, followed by entropy encoding \cite{witten1987arithmetic} to convert $\boldsymbol{\bar y}$ into bit sequence. This bit sequence will be fed into entropy decoding to losslessly reconstruct $\boldsymbol{\bar y}$, and another nonlinear synthesis transform DNN module $g_s$ uses $\boldsymbol{\bar y}$ to reconstruct the decoded data $\boldsymbol{\hat x}$. $g_a$ and $g_s$ are jointly optimized under the rate-distortion constraint. In communication systems, the above source compressive coding paradigm relies heavily on advanced channel coding and signal processing techniques to ensure the transmitted bit sequence to be losslessly recovered. This separation-based approach has been employed in many current communication systems, as the binary representations of various source data can be seamlessly transmitted over arbitrary wireless channels by changing the underlying channel code.

However, with increasing demands on low-latency wireless data delivery applications such as extended reality (XR), the limits of the separation-based design begin to emerge. Current wireless data transmission systems suffer from time-varying channel conditions, in which case the separation-based design leads to significant \emph{cliff-effect} when the channel condition is below the level anticipated by the channel code \cite{DJSCC}. Furthermore, the widely-used entropy coding is quite sensitive to the variational estimate of the marginal distribution of the source latent representation. Small perturbations on this marginal can lead to the catastrophic error propagation in entropy decoding \cite{rissanen1981universal}. In practice, the small perturbation is often caused by the floating point round-off error \cite{balle2018integer}. This round-off operation depends heavily on hardware and software platforms, and in various data compression applications, the transceiver may employ different platforms as stated in \cite{balle2018integer}. As a result, this non-determinism issue in transmitter vs. receiver will lead to severe performance degradation.

\begin{figure}[t]
	\centering{\includegraphics[scale=0.42]{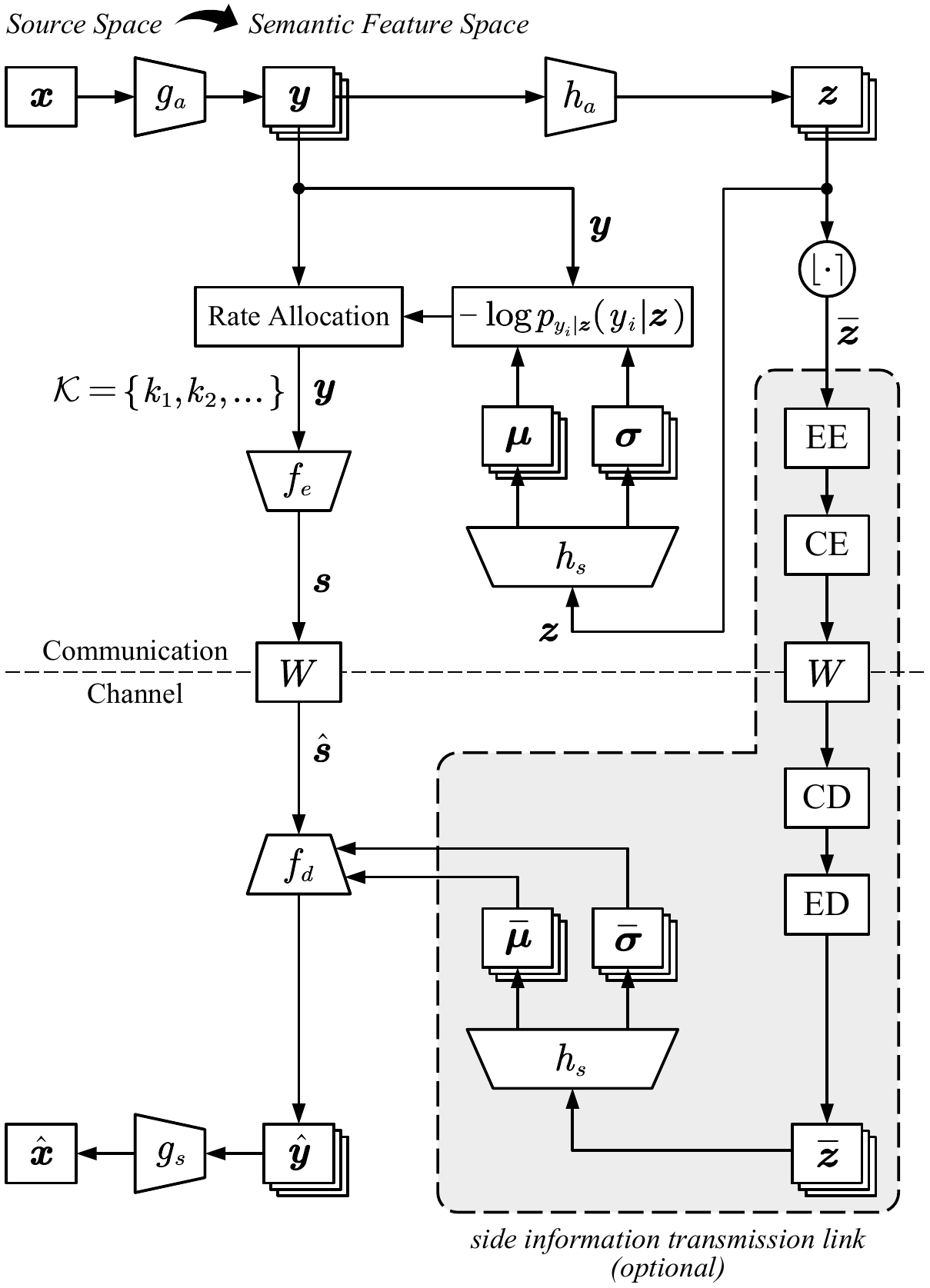}}
	\caption{The whole NTSCC architecture for semantic communications. The transmitter should know the entropy model on $\boldsymbol{\bar z}$ to entropy encode (EE) and channel encode (CE) it, that is modeled as a non-parametric factorized density conditioned on $\boldsymbol{\psi}$ as \eqref{eq_entropy_model_z}. The receiver begins with channel decoding (CD) and entropy decoding (ED) to recover the side information $\boldsymbol{\bar z}$, and then uses it to decode $\boldsymbol{\hat y}$. In addition, the side information $\boldsymbol{\bar z}$ is not necessary for the receiver as analyzed in \cite{dai2022nonlinear}. If $\boldsymbol{\bar z}$ is not transmitted, the decoding performance shows some degradation while the bandwidth cost is also reduced. On the whole, the system bandwidth cost-distortion performance is comparable.}\label{Fig_ntscc_arch}
	\vspace{0em}
\end{figure}

To address the above issues, our idea in NTSCC \cite{dai2022nonlinear} is to \emph{replace quantization and entropy coding} by integrating source coding and channel coding as a trained DNN, resulting in deep JSCC to transmit the latent representation $\boldsymbol{y}$ directly. We have achieved better end-to-end transmission performance, and the system is robust to unpredictive wireless channels. The whole procedure of NTSCC is depicted in Fig. \ref{Fig_ntscc_arch}. The latent code $\boldsymbol{y}$ is fed into both the analysis transform $h_a$ and the deep JSCC encoder $f_e$. On the one hand, $h_a$ summarizes the distribution of mean values and standard derivations of $\boldsymbol{y}$ in the hyperprior $\boldsymbol{z}$. The transmitter utilizes $\boldsymbol{z}$ to estimate the mean vector $\boldsymbol{\mu}$ and the standard derivation vector $\boldsymbol{\sigma}$, and use them to determine the bandwidth to transmit the latent representation. On the other hand, $f_e$ encodes $\boldsymbol{y}$ as the channel-input sequence $\boldsymbol{s} \in \mathbb{R}^k$, and the received sequence is ${\boldsymbol{\hat s}} = W( \boldsymbol{s} )$, whose transition probability is ${{p_{{\boldsymbol{\hat s}}| {\boldsymbol{s}} }}( {{\boldsymbol{\hat s}}| \boldsymbol{s} } )}$. In this paper, we consider the general fading channel model such that the transfer function is ${\boldsymbol{\hat s}} = W( \boldsymbol{s}|\boldsymbol{h} ) = \boldsymbol{h} \odot \boldsymbol{s} + \boldsymbol{n}$ where
$\odot$ is the element-wise product, $\boldsymbol{h}$ denotes the CSI vector, and each component of the noise vector $\boldsymbol{n}$ is independently sampled from a Gaussian distribution, i.e., $\boldsymbol{n} \sim p_{\boldsymbol{n}} \triangleq \mathcal{N}(\boldsymbol{0}, {\sigma_n^2}{\boldsymbol{I}})$, where ${\sigma_n^2}$ is the noise power. At the receiver, ${\boldsymbol{\hat s}}$ is further fed into the deep JSCC decoder $f_d$ to reconstruct the latent representation $\boldsymbol{\hat y}$, which is further used by the nonlinear synthesis transform $g_s$ to recover the source data $\boldsymbol{\hat x}$. The whole procedure of NTSCC system is
\begin{equation}\label{eq_trans_process}
\begin{aligned}
& {{\boldsymbol{x}}} \xrightarrow{{{g_a}( \cdot ;{\boldsymbol{\phi}}_g)}} {{\boldsymbol{y}}}
\xrightarrow{{{f_e}( \cdot ;{\boldsymbol{\phi}}_f)}} {{\boldsymbol{s}}}
\xrightarrow{{{W}( \cdot|\boldsymbol{h} )}} {{\boldsymbol{\hat s}}}
\xrightarrow{{{f_d}( \cdot ;{\boldsymbol{\theta}}_f)}} {{\boldsymbol{\hat y}}}
\xrightarrow{{{g_s}( \cdot ;{\boldsymbol{\theta}}_g)}} {{\boldsymbol{\hat x}}} \\
& \text{with the latent prior~} {{\boldsymbol{y}}} \xrightarrow{{{h_a}( \cdot ;{\boldsymbol{\phi}}_h)}} {{\boldsymbol{z}}}
\xrightarrow{{{h_s}( \cdot ;{\boldsymbol{\theta}}_h)}} \left\{ {{\boldsymbol{\mu}},{\boldsymbol{\sigma}}} \right\},
\end{aligned}
\end{equation}
where $(\boldsymbol{\phi}, \boldsymbol{\theta}) = (\boldsymbol{\phi}_g,\boldsymbol{\phi}_h,\boldsymbol{\phi}_f,\boldsymbol{\theta}_g,\boldsymbol{\theta}_h,\boldsymbol{\theta}_{f})$ encapsulate learnable DNN parameters of each function. The system efficiency is measured by the \emph{channel bandwidth ratio (CBR)} $\rho = k/m$.

The key idea of NTSCC lies in variable-length deep JSCC guided by the latent prior on the semantic feature space. The latent prior $p_{\boldsymbol{y} | \boldsymbol{z}} (\boldsymbol{y} | \boldsymbol{ z})$ is obtained as
\begin{equation}\label{eq_ntscc_z_to_y}
\begin{aligned}
  p_{\boldsymbol{y} | \boldsymbol{z}} (\boldsymbol{y} | \boldsymbol{ z}) = & \prod_i \underbrace{\left( \mathcal{N}(y_i|{{\mu}}_i,{{\sigma}}_i^2) * \mathcal{U}(-\frac{1}{2},\frac{1}{2}) \right)}_{p_{y_i | \boldsymbol{z}}} ({ y}_i) \\
  ~ & \text{with~} (\boldsymbol{ \mu},\boldsymbol{ \sigma}) = {h_s}(\boldsymbol{ z}; \boldsymbol{\theta}_h),
\end{aligned}
\end{equation}
where the convolutional operation ``$*$'' with a standard uniform distribution is used to match the prior to the marginal such that the estimated rate $-\log{p_{\boldsymbol{y} | \boldsymbol{z}} (\boldsymbol{y} | \boldsymbol{ z})}$ is non-negative. The hyperprior $\boldsymbol{z}$ is usually transmitted over the digital link as side information due to its small cost, where the quantization $\lfloor \cdot \rceil$ (rounding to integers) is needed as marked in Fig. \ref{Fig_ntscc_arch}. A uniformly-noised proxy $\boldsymbol{\tilde z} = \boldsymbol{z} + \boldsymbol{o}$ is used to replace the quantized representation $\boldsymbol{\bar y} = \lfloor \boldsymbol{y} \rceil$ during model training \cite{balle2018}, where $o_j$ is sampled from $\mathcal{U}(-\frac{1}{2},\frac{1}{2})$. The probability of hyperprior $\boldsymbol{\tilde z}$ is calculated on the fully factorized density $p_{\boldsymbol{z}} = \prod\nolimits_j p_{{z}_j}$ as
\begin{equation}\label{eq_entropy_model_z}
  p_{\boldsymbol{z}} (\boldsymbol{\tilde z}) = \prod_j  \underbrace{\left( p_{{z}_j | \boldsymbol{\psi}^{(j)}} ({z}_j | \boldsymbol{\psi}^{(j)}) * \mathcal{U}(-\frac{1}{2},\frac{1}{2}) \right)}_{p_{{z}_j}} ({\tilde z}_j),
\end{equation}
where $\boldsymbol{\psi}^{(j)}$ encapsulates all the parameters of $p_{{z}_j | \boldsymbol{\psi}^{(j)}}$.

The optimizing problem of NTSCC is formulated following the variational inference context \cite{kingma2013auto}, the posterior distribution $p_{\boldsymbol{\hat s}, {\boldsymbol{\tilde z}} | \boldsymbol{x}}$ is approximated using the variational density $q_{\boldsymbol{\hat s}, {\boldsymbol{\tilde z}} | \boldsymbol{x}}$ by minimizing their Kullback-Leibler (KL) divergence over the data distribution $p_{\boldsymbol{x}}$ and the CSI distribution $p_{\boldsymbol{h}}$ as the equation (11) in \cite{dai2022nonlinear}. Accordingly, the optimization of NTSCC system can be formally converted to the minimization of the expected channel bandwidth cost, as well as the expected distortion of the reconstructed data versus the original, which leads to the optimization of the following R-D trade-off,
\begin{equation}\label{eq_expect_loss_func}
\begin{aligned}
  & \mathcal{L}_{\text{R-D}}(\boldsymbol{\phi}, \boldsymbol{\theta}, \boldsymbol{\psi}) = \mathbb{E}_{\boldsymbol{x}\sim p_{\boldsymbol{x}}}\mathbb{E}_{\boldsymbol{h}\sim p_{\boldsymbol{h}}} D_{\rm{KL}}(q_{\boldsymbol{\hat s},\boldsymbol{\tilde z} | \boldsymbol{x}} \| p_{\boldsymbol{\hat s},\boldsymbol{\tilde z} | \boldsymbol{x}} ) \Leftrightarrow \mathbb{E}_{\boldsymbol{x}\sim p_{\boldsymbol{x}}}  \\
  & \mathbb{E}_{\boldsymbol{h}\sim p_{\boldsymbol{h}}} \Big( \lambda \underbrace{\big( -{\eta_y} \log{p_{\boldsymbol{ y}|\boldsymbol{ z}}(\boldsymbol{ y}|\boldsymbol{ z})} - {\eta_z}{\log{p_{\boldsymbol{z}}(\boldsymbol{\tilde z})}} \big)}_{\text{data stream channel bandwidth cost:~}R} + \underbrace{d(\boldsymbol{x},\boldsymbol{\hat{x}})}_{\text{distortion:~}D}\Big),
\end{aligned}
\end{equation}
where the Lagrange multiplier $\lambda$ on the total channel bandwidth cost determines the trade-off between the data stream bandwidth cost $R$ and the end-to-end distortion $D$. The scaling factors $\eta_y$ and $\eta_z$ control the relation between the estimated entropy and the allocated channel bandwidth, which are tied with the source-channel codec capability and the wireless channel state. A larger $\eta_y$ indicates a better performance on deep JSCC codec $f_e$ and $f_d$, but incurs more channel bandwidth cost. Accordingly, $\eta_y$ can be adjusted as a hyperparameter to control the system R-D trade-off. $\eta_z$ is not adjusted manually since explicit entropy coding and LDPC coding are selected to transmit the side information.

In practice, each embedding $y_i$ is a $c$-dimensional feature vector. The learned entropy model $-\log p_{{ y}_i|\boldsymbol{ z}}({ y}_i|\boldsymbol{ z})$ indicates the summation of entropy along $c$ dimensions of $y_i$, thus, the information density distribution of $\boldsymbol{y}$ is captured. Accordingly, the bandwidth cost, such as the number of OFDM subcarriers, ${\bar k}_{i}$ for transmitting $y_i$ can be determined as
\begin{equation}\label{eq_channel_bandwidth_cost_cal}
  {\bar k}_{i} = Q({k}_{i}) = Q\Big( \underbrace{-{\eta_y} \log{p_{y_i|\boldsymbol{ z}}(y_i|\boldsymbol{ z})}}_{k_i} \Big),
\end{equation}
where the learned entropy model $p_{{ y}_i|\boldsymbol{ z}}$ follows \eqref{eq_ntscc_z_to_y}, $Q$ denotes a scalar quantization whose range includes $2^{q}$ ($q = 1,2,\dots$) integers, and the quantization value set $\mathcal{V} = \{ v_1, v_2, \dots, v_{2^{q}} \}$ is related to the scaling factor $\eta_y$ and the Lagrange multiplier $\lambda$. Hence, the predetermined $q$ bits should be transmitted as extra side information to inform the receiver which bandwidth is allocated to every embedding $y_i$. To adaptively map $y_i$ to a ${\bar k}_{i}$-dimensional channel-input vector $s_i$, the dynamic neural network structure \cite{han2021dynamic} is introduced into Transformers \cite{vaswani2017attention} to realize the deep JSCC codec $f_e$ and $f_d$ \cite{dai2022nonlinear}.

\subsection{Motivation of Online Learned Adaptive NTSCC}

In existing works, the loss function $\mathcal{L}_{\text{R-D}}$ in \eqref{eq_expect_loss_func} is optimized over a corpus of source data samples (such as a large amount of images) and channel states in order to find optimal codec function parameters $\boldsymbol{\Sigma} = (\boldsymbol{\phi}, \boldsymbol{\theta}, \boldsymbol{\psi})$. Although the models have been trained over a large corpus of source and channel samples for finding out what should be ideally optimal codec functions $(g_a$, $g_s$, $h_a$, $h_s$, $f_e$, $f_d)$ over the whole data set and ergodic channel responses, we will show the codec functions can still be improved for each single data sample and instant CSI. This suboptimality of NTSCC model can be interpreted from the \emph{inference suboptimality} of VAE \cite{cremer2018inference} as follows: the mismatch between the true and approximate posterior. It has been proven that the inference gap includes two components: the \emph{approximation gap} and the \emph{amortization gap}. This approximation gap comes from the inability of the variational distribution family to exactly match the true posterior, and the amortization gap refers to the difference caused by amortizing the variational parameters over the entire training set, instead of optimizing for each training example individually.

\begin{figure}[t]
	\centering{\includegraphics[scale=0.5]{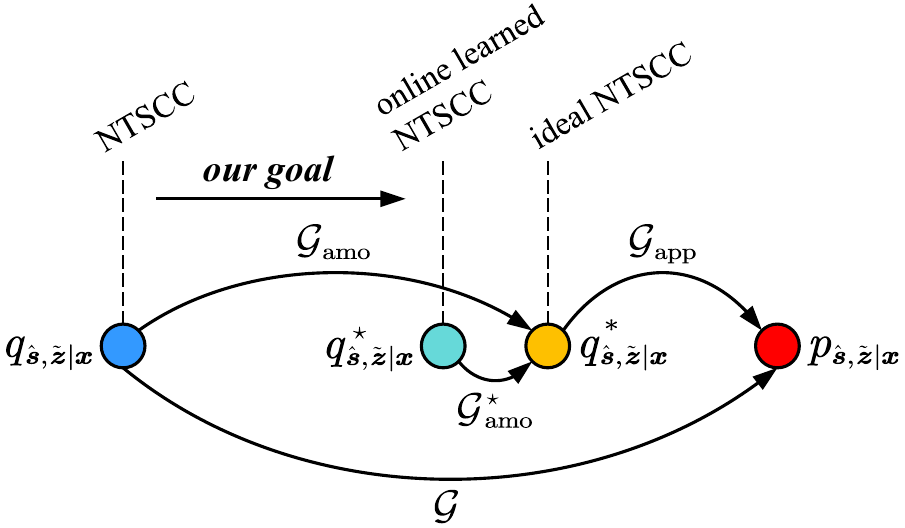}}
	\caption{The relation of gaps in model inference. By overfitting the source and channel, we can upgrade the standard NTSCC system to the online learned NTSCC system, thus making $\mathcal{G}_{\text{amo}}^{\star} \ll \mathcal{G}_{\text{amo}}$.}\label{Fig_motivation}
	\vspace{-1em}
\end{figure}

Specifically, in our NTSCC system, given the source data sample $\boldsymbol{x}$ and the instant CSI vector $\boldsymbol{h}$, the inference gap $\mathcal{G}$ is
\begin{equation}\label{eq_inference_gap}
  \mathcal{G} = D_{\rm{KL}}(q_{\boldsymbol{\hat s},\boldsymbol{\tilde z} | \boldsymbol{x}} \| p_{\boldsymbol{\hat s},\boldsymbol{\tilde z} | \boldsymbol{x}} ),
\end{equation}
where $q_{\boldsymbol{\hat s},\boldsymbol{\tilde z} | \boldsymbol{x}}$ is derived using the parameters $(\boldsymbol{\phi}, \boldsymbol{\theta}, \boldsymbol{\psi})$ learned under the entire training set, e.g.,
\begin{equation}
  q_{\boldsymbol{\hat s},\boldsymbol{\tilde z} | \boldsymbol{x}} = \arg{\mathop {\min }\limits_{q \in \mathcal{Q}}}\mathbb{E}_{\boldsymbol{x}\sim p_{\boldsymbol{x}}}\mathbb{E}_{\boldsymbol{h}\sim p_{\boldsymbol{h}}} D_{\rm{KL}}(q_{\boldsymbol{\hat s},\boldsymbol{\tilde z} | \boldsymbol{x}} \| p_{\boldsymbol{\hat s},\boldsymbol{\tilde z} | \boldsymbol{x}} ).
\end{equation}
However, the optimal $q_{\boldsymbol{\hat s},\boldsymbol{\tilde z} | \boldsymbol{x}}^{*}$ should be derived under the given $\boldsymbol{x}$ and $\boldsymbol{h}$ as
\begin{equation}
  q_{\boldsymbol{\hat s},\boldsymbol{\tilde z} | \boldsymbol{x}}^{*} = \arg{\mathop {\min }\limits_{q \in \mathcal{Q}}} D_{\rm{KL}}(q_{\boldsymbol{\hat s},\boldsymbol{\tilde z} | \boldsymbol{x}} \| p_{\boldsymbol{\hat s},\boldsymbol{\tilde z} | \boldsymbol{x}} ),
\end{equation}
which corresponds to the optimal parameters $(\boldsymbol{\phi}^{*}, \boldsymbol{\theta}^{*}, \boldsymbol{\psi}^{*})$. As illustrated in Fig. \ref{Fig_motivation}, it can be derived by
\begin{equation}\label{eq_inference_gap_derive}
\begin{aligned}
  & \mathcal{G} =  D_{\rm{KL}}(q_{\boldsymbol{\hat s},\boldsymbol{\tilde z} | \boldsymbol{x}} \| p_{\boldsymbol{\hat s},\boldsymbol{\tilde z} | \boldsymbol{x}} ) = \underbrace{D_{\rm{KL}}(q_{\boldsymbol{\hat s},\boldsymbol{\tilde z} | \boldsymbol{x}}^{*} \| p_{\boldsymbol{\hat s},\boldsymbol{\tilde z} | \boldsymbol{x}} )}_{\text{approximation gap:~}\mathcal{G}_{\text{app}}} + \\
  & \underbrace{D_{\rm{KL}}(q_{\boldsymbol{\hat s},\boldsymbol{\tilde z} | \boldsymbol{x}} \| p_{\boldsymbol{\hat s},\boldsymbol{\tilde z} | \boldsymbol{x}} ) - D_{\rm{KL}}(q_{\boldsymbol{\hat s},\boldsymbol{\tilde z} | \boldsymbol{x}}^{*} \| p_{\boldsymbol{\hat s},\boldsymbol{\tilde z} | \boldsymbol{x}} )}_{\text{amortization gap:~}\mathcal{G}_{\text{amo}}} = \mathcal{G}_{\text{app}} + \mathcal{G}_{\text{amo}}.
\end{aligned}
\end{equation}
This indicates that the amortized posterior $q$ over the entire source dataset and ergodic CSI still incurs the amortization gap $\mathcal{G}_{\text{amo}}$ under the instant $\boldsymbol{x}$ and $\boldsymbol{h}$. The above analysis demonstrates that a neural-network-based semantic communication model is trained on the entire dataset and CSI set with the target of achieving the best end-to-end R-D performance on test data and CSI, i.e., we ideally expect $\mathcal{G}_{\text{amo}} = 0 \Leftrightarrow \mathcal{G} = \mathcal{G}_{\text{app}}$. However, due to limited model capacity, optimization difficulties, and insufficient data and CSI, the model cannot in general achieve this goal. When the source data distribution or channel model differs from that in the training phase, model generalization will not be guaranteed even with the infinite data and model capacity, and perfect optimization.

We however note that a convenient feature of neural wireless data transmission is a model or semantic latent representation that can be easily finetuned on new data and CSI. A model can for instance or domain be trained after deployment. Inspired by this, as illustrated in Fig. \ref{Fig_motivation}, our goal is closing the amortization gap $\mathcal{G}_{\text{amo}}$ by overfitting the source and channel such that the resulting posterior $q_{\boldsymbol{\hat s},\boldsymbol{\tilde z} | \boldsymbol{x}}^{\star}$ of our online learned NTSCC model can approach the optimal $q_{\boldsymbol{\hat s},\boldsymbol{\tilde z} | \boldsymbol{x}}^{*}$ at every test instance. Thus the new amortization gap $\mathcal{G}_{\text{amo}}^{\star}$ is much smaller than $\mathcal{G}_{\text{amo}}$. As such, we are effectively trying to solve the following optimization problem, given the instant $\boldsymbol{x}$ and $\boldsymbol{h}$,
\begin{equation}\label{eq_instant_loss_func}
\begin{aligned}
  & (\boldsymbol{\phi}^{*}, \boldsymbol{\theta}^{*}, \boldsymbol{\psi}^{*}) = \arg{\mathop {\min }\limits_{\boldsymbol{\phi}, \boldsymbol{\theta}, \boldsymbol{\psi}}}D_{\rm{KL}}(q_{\boldsymbol{\hat s},\boldsymbol{\tilde z} | \boldsymbol{x}} \| p_{\boldsymbol{\hat s},\boldsymbol{\tilde z} | \boldsymbol{x}} ) \\
  & = \arg{\mathop {\min }\limits_{\boldsymbol{\phi}, \boldsymbol{\theta}, \boldsymbol{\psi}}} \mathcal{L}_{\text{R-D}}(\boldsymbol{\phi}, \boldsymbol{\theta}, \boldsymbol{\psi}, \boldsymbol{x}, \boldsymbol{h}) =  \arg{\mathop {\min }\limits_{\boldsymbol{\phi}, \boldsymbol{\theta}, \boldsymbol{\psi}}} \\
  &  \Big( \lambda \underbrace{\big( -{\eta_y} \log{p_{\boldsymbol{ y}|\boldsymbol{ z}}(\boldsymbol{ y}|\boldsymbol{ z})} - {\eta_z}{\log{p_{\boldsymbol{ z}}(\boldsymbol{\tilde z})}} \big)}_{\text{data stream bandwidth cost:~}R} + \underbrace{d(\boldsymbol{x},\boldsymbol{\hat{x}})}_{\text{distortion:~}D} \Big),
\end{aligned}
\end{equation}
where $\mathcal{L}_{\text{R-D}}(\boldsymbol{\phi}, \boldsymbol{\theta}, \boldsymbol{\psi}, \boldsymbol{x}, \boldsymbol{h})$ is the transmission R-D objective for a particular $\boldsymbol{x}$ and $\boldsymbol{h}$.

\section{Overfitting the Source}\label{section_source_overfit}

In this section, we present two methods to overfit a single source data instance $\boldsymbol{x}$ (e.g. an image) or data from a specific domain $\mathcal{X}^\prime$ (e.g. a set of I-frames from video sequences in the same scene). They stem from two different ideas:
\begin{itemize}
  \item \emph{Transmitter adaptation (Tx-adapt):} Given an instance $\boldsymbol{x}$ or a group of domain samples from $\mathcal{X}^\prime$, online update ($g_a$, $f_e$) or ($\boldsymbol{y}$, $\boldsymbol{s}$) using gradient descent based on the off-the-shelf pre-trained model.

  \item \emph{Transceiver adaptation (TxRx-adapt):} Given a group of domain samples from $\mathcal{X}^\prime$, online update ($g_a$, $f_e$, $f_d$, $g_s$) using gradient descent based on the off-the-shelf pre-trained model.
\end{itemize}

\begin{figure}[t]
	\centering{\includegraphics[scale=0.29]{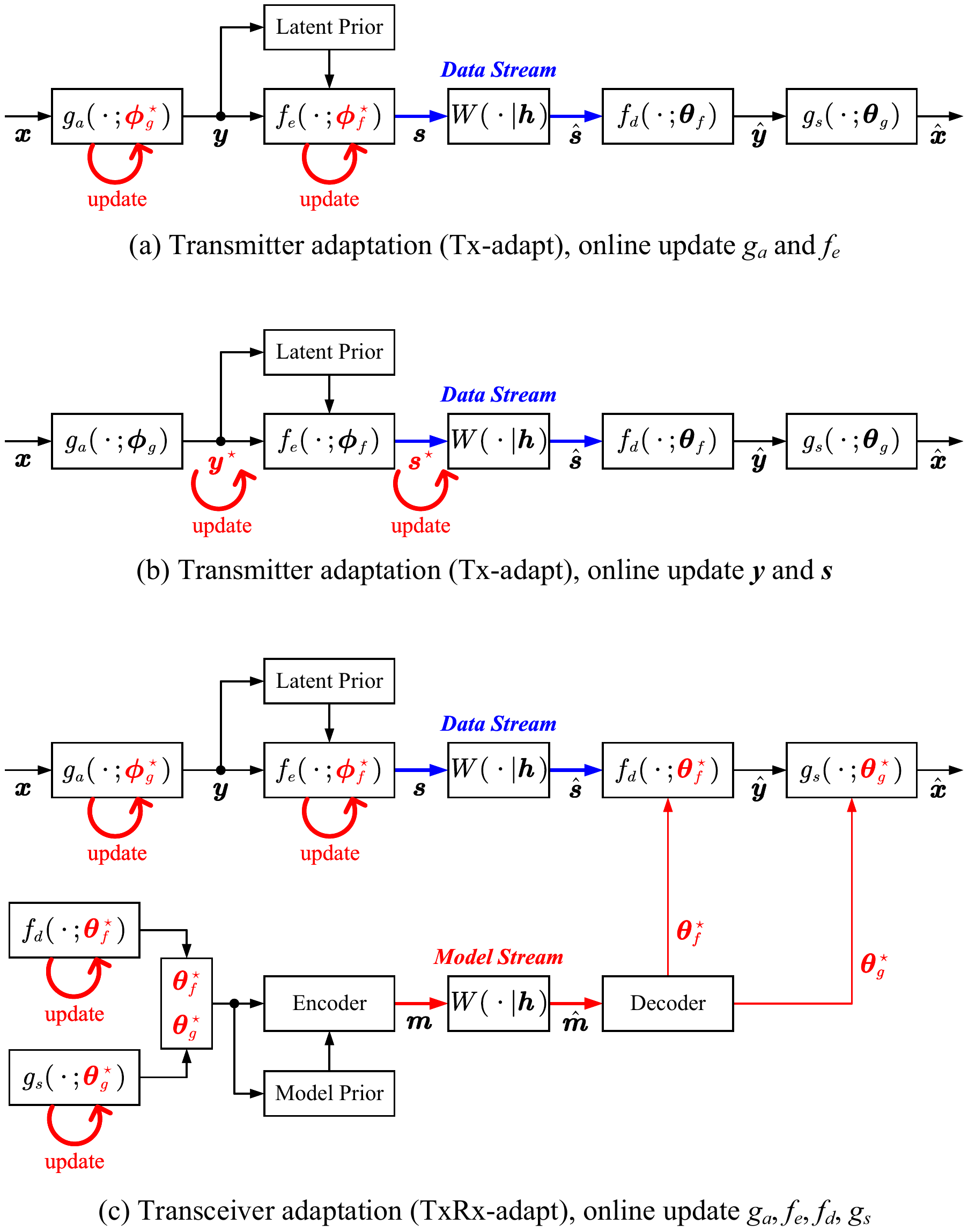}}
	\caption{Key ideas of overfitting the source in the online learned NTSCC.}\label{Fig4}
	\vspace{0em}
\end{figure}

Fig. \ref{Fig4} presents the key idea of our proposed methods, where red lines mark the updated ingredients during model inference phase. In the following subsections, the instant CSI $\boldsymbol{h}$ will be sampled from a specific channel model $\mathcal{H}^\prime$, we present how to overfit the source data instance or a specific domain.

Our goal is adapting the baseline model ${\boldsymbol{\Sigma}}_{(\mathcal{X},\mathcal{H})}$ learned on the training dataset $\mathcal{X}$ and channel state domain $\mathcal{H}$ to the given specific source domain $\mathcal{X}^\prime$ and channel domain $\mathcal{H}^\prime$. In general, $\mathcal{X}^\prime$ can be a subset of $\mathcal{X}$ or a related set to $\mathcal{X}$, and $\mathcal{H}^\prime$ is a partial scene of $\mathcal{H}$, e.g., the CSI set of a local area. In this section, we focus on solving the following problems,
\begin{subequations}\label{eq_source_overfit_goal}
\begin{equation}
    \text{\emph{source instance adaptation:}~}{\boldsymbol{\Sigma}}_{(\mathcal{X},\mathcal{H})} \Rightarrow {\boldsymbol{\Sigma}}_{(\boldsymbol{x},\mathcal{H}^\prime)},
\end{equation}
\begin{equation}
    \text{\emph{source domain adaptation:}~}{\boldsymbol{\Sigma}}_{(\mathcal{X},\mathcal{H})} \Rightarrow {\boldsymbol{\Sigma}}_{(\mathcal{X}^\prime,\mathcal{H}^\prime)}.
\end{equation}
\end{subequations}
Our models ${\boldsymbol{\Sigma}}_{(\boldsymbol{x},\mathcal{H}^\prime)}$ and ${\boldsymbol{\Sigma}}_{(\mathcal{X}^\prime,\mathcal{H}^\prime)}$ are both incidentally adapted to a local channel domain $\mathcal{H}^\prime$ since we assume the exact CSI is only available at the receiver (CSIR) instead of the transmitter. In this case, the transmitter cannot obtain the exact CSI instance $\boldsymbol{h} \in \mathcal{H}^\prime$ in real time. Thus, in our source overfitting algorithms described following, the CSI vector $\boldsymbol{h}$ is sampled from the specific channel domain $\mathcal{H}^\prime$ known at the transmitter. As for how to adapt our model to the exact CSI vector, it will be discussed in the next section.

\subsection{Transmitter Adaptation}

Given the source sample $\boldsymbol{x}$ and the baseline model $(\boldsymbol{\phi}, \boldsymbol{\theta}, \boldsymbol{\psi})$ learned on the entire dataset, our goal is adapting the transmitter model parameters $(\boldsymbol{\phi}_g, \boldsymbol{\phi}_f)$ or the latent code and channel-input codeword $(\boldsymbol{y}, \boldsymbol{s})$ to every single data sample.

The key benefit of transmitter adaptation is achieving an improved end-to-end transmission performance while keeping the predictive model fixed such that the computing time at the receiver stays unchanged. The refined nonlinear analysis transform $g_a$ or the latent code $\boldsymbol{y}$ provides a more compact semantic representation of the source sample $\boldsymbol{x}$, leading to the superior end-to-end R-D performance. As a trade-off, the transmitter incurs additional encoding time and computational expense, while the decoding delay stays unchanged.

This adaptation mode is suitable for situations where sufficient transmitter side computational power is available (e.g., the media server, often located in a relatively well-provisioned facility, such as the cloud), and the encoding delay is not a pressing issue. As a typical case, existing high-resolution image/video wireless delivery depends critically on the bandwidth resource \cite{jiang2014improving}, and video content is known in advance at the capable server side. In this case, the transmitter adaptation paradigm can significantly enhance user quality of experience by utilizing server computation.

For model adaptation, we aim at effectively solving the following optimization problem. During the model inference time, for a single data instance $\boldsymbol{x}$:
\begin{equation}\label{eq_target_Tx_model_adp}
  \begin{aligned}
  (\boldsymbol{\phi}_g^{\star}, \boldsymbol{\phi}_f^{\star}) = \arg{\mathop {\min }\limits_{\boldsymbol{\phi}_g, \boldsymbol{\phi}_f}}  \mathcal{L}_{\text{R-D}}(\boldsymbol{\phi}, \boldsymbol{\theta}, \boldsymbol{\psi}, \boldsymbol{x}, \boldsymbol{h}),
\end{aligned}
\end{equation}
where we adapt $g_a$ and $f_e$ while fixing the entropy model $h_a$ and $h_s$. This design aims to reduce the model updating complexity. In addition, if the side information $\boldsymbol{\bar{z}}$ is transmitted, our method ensures that no model updates have to be transmitted to update $h_s$ at the receiver. Experiments can verify that our simplified method achieves comparable performance as that of finetuning the whole transmitter model $\boldsymbol{\phi} = (\boldsymbol{\phi}_g, \boldsymbol{\phi}_f, \boldsymbol{\phi}_h)$. In this work, we solve this problem \eqref{eq_target_Tx_model_adp} in an iterative procedure as that in \cite{campos2019content}. We apply gradient descent on $\boldsymbol{\phi}_g$ and $\boldsymbol{\phi}_f$ to update the nonlinear analysis transform $g_a$ and the deep JSCC encoder $f_e$. The iterative parameter updating procedures are
\begin{subequations}\label{eq_Tx_model_adp_iterative}
\begin{equation}\label{eq_Tx_model_adp_iterative_a}
  \boldsymbol{\phi}_g^{(t)} = \boldsymbol{\phi}_g^{(t-1)} - \gamma \nabla_{\boldsymbol{\phi}_g} \mathcal{L}_{\text{R-D}}(\boldsymbol{\phi}, \boldsymbol{\theta}, \boldsymbol{\psi}, \boldsymbol{x}, \boldsymbol{h}),
\end{equation}
\begin{equation}\label{eq_Tx_model_adp_iterative_b}
   \boldsymbol{\phi}_f^{(t)} = \boldsymbol{\phi}_f^{(t-1)} - \gamma \nabla_{\boldsymbol{\phi}_f} \mathcal{L}_{\text{R-D}}(\boldsymbol{\phi}, \boldsymbol{\theta}, \boldsymbol{\psi}, \boldsymbol{x}, \boldsymbol{h}),
\end{equation}
\end{subequations}
where $\gamma$ denotes the learning rate. The final pipeline of Tx model adaptation is described in Algorithm \ref{alg_tx_adp_model}.

\begin{algorithm}[t]
\setlength{\abovecaptionskip}{0cm}
\setlength{\belowcaptionskip}{-0cm}

\caption{Tx Model Online Adaptation}\label{alg_tx_adp_model}

\KwIn{Baseline model parameters $(\boldsymbol{\phi}, \boldsymbol{\theta}, \boldsymbol{\psi})$ trained on training set, the data sample to be transmitted $\boldsymbol{x}$, the CSI vector $\boldsymbol{h}$.}

\KwOut{The updated models $\boldsymbol{\phi}_g^{\star}$ and $\boldsymbol{\phi}_f^{\star}$.}

\SetKwBlock{BeginR}{procedure~\textnormal{\textsc{OnlineUpdateModels}$(\boldsymbol{\phi}, \boldsymbol{\theta}, \boldsymbol{\psi}, \boldsymbol{x}, \boldsymbol{h})$}}{}

\BeginR{
Initialize model parameters: $\boldsymbol{\phi}_g^{(0)} \leftarrow \boldsymbol{\phi}_g$, $\boldsymbol{\phi}_f^{(0)} \leftarrow \boldsymbol{\phi}_f$\;

\For{$t = 1,2,\dots,T_{\max}$}
{
    Forward pass: ${{\boldsymbol{x}}} \xrightarrow{{{g_a}( \cdot ;{\boldsymbol{\phi}}_g^{(t-1)})}} {{\boldsymbol{y}}}
\xrightarrow{{{f_e}( \cdot ;{\boldsymbol{\phi}}_f^{(t-1)})}} {{\boldsymbol{s}}}
\xrightarrow{{{W}( \cdot|\boldsymbol{h})}} {{\boldsymbol{\hat s}}}
\xrightarrow{{{f_d}( \cdot ;{\boldsymbol{\theta}}_f)}} {{\boldsymbol{\hat y}}}
\xrightarrow{{{g_s}( \cdot ;{\boldsymbol{\theta}}_g)}} {{\boldsymbol{\hat x}}}$ with the latent prior ${{\boldsymbol{y}}} \xrightarrow{{{h_a}( \cdot ;{\boldsymbol{\phi}}_h)}} {{\boldsymbol{z}}}
\xrightarrow{{{h_s}( \cdot ;{\boldsymbol{\theta}}_h)}} \left\{ {{\boldsymbol{\mu}},{\boldsymbol{\sigma}}} \right\}$\;

    Compute loss $\mathcal{L}_{\text{R-D}}$ according to \eqref{eq_instant_loss_func}\;

    Update $\boldsymbol{\phi}_g^{(t)}$ and $\boldsymbol{\phi}_f^{(t)}$ using gradients $\nabla_{\boldsymbol{\phi}_g} \mathcal{L}_{\text{R-D}}$ and $\nabla_{\boldsymbol{\phi}_f} \mathcal{L}_{\text{R-D}}$ as \eqref{eq_Tx_model_adp_iterative}\;

}

\Return Updated models $\boldsymbol{\phi}_g^{\star} \leftarrow \boldsymbol{\phi}_g^{(T_{\max})}$ and $\boldsymbol{\phi}_f^{\star} \leftarrow \boldsymbol{\phi}_f^{(T_{\max})}$.
}

\end{algorithm}

Apparently, instance adaptation is an extreme case of domain adaptation, which provides better performance with the expense of more encoding time. In practice, we pursue for more efficient online adaptation, thus the Tx model adaptation in Algorithm \ref{alg_tx_adp_model} is often invoked only for several spare source samples from the same domain, e.g., a set of I-frames from a single video whose content locates in the same scene. We also note that since the decoder parameters remain unchanged such that the performance improvements are limited. A more effective transceiver adaptation mode tailored for domain adaptation will be explored in the subsequent subsection.

Next, we introduce a more lightweight method to adapt the latent representation $\boldsymbol{y}$ and the deep JSCC codeword $\boldsymbol{s}$ for instance adaptation only. This code adaptation method aims to find more compact semantic representations directly without changing the encoder/decoder parameters. We apply gradient descent on $\boldsymbol{y}$ and $\boldsymbol{s}$ to update the semantic latent representation and the deep JSCC codeword, respectively. A special note is that the iterative updating procedure on these two terms should be executed sequentially because $\boldsymbol{s}$ is generated from $\boldsymbol{y}$ by using the function $f_e$. The iterative updating procedure can be written as
\begin{equation}\label{eq_Tx_code_adp_iterative_a}
\begin{aligned}
  & \boldsymbol{y}^{(t)} = \boldsymbol{y}^{(t-1)} - \gamma \nabla_{\boldsymbol{y}} \mathcal{L}_{\text{R-D}}(\boldsymbol{\phi}, \boldsymbol{\theta}, \boldsymbol{\psi}, \boldsymbol{x}, \boldsymbol{h}),\\
  & \text{with~} t = 1,2,\dots,Y_{\max}.
\end{aligned}
\end{equation}
After $t$ reaches $Y_{\max}$, the procedure turns to update $\boldsymbol{s}$, i.e.,
\begin{equation}\label{eq_Tx_code_adp_iterative_b}
\begin{aligned}
  & \boldsymbol{s}^{(t)} = \boldsymbol{s}^{(t-1)} - \gamma \nabla_{\boldsymbol{s}} \mathcal{L}_{\text{R-D}}(\boldsymbol{\phi}, \boldsymbol{\theta}, \boldsymbol{\psi}, \boldsymbol{x}, \boldsymbol{h}),\\
  & \text{with~} t = 1,2,\dots,S_{\max}.
\end{aligned}
\end{equation}
The total number of updating steps is $T_{\max} = Y_{\max} + S_{\max}$. The final pipeline of Tx code adaptation is described by Algorithm \ref{alg_tx_adp_code}. This latent representation and codeword adaptation technique reduces the number of updated parameters for lower complexity and decreased GPU peak memory, but it is only applicable for instance adaptation, not for domain adaptation.

\begin{algorithm}[t]
\setlength{\abovecaptionskip}{0cm}
\setlength{\belowcaptionskip}{-0cm}

\caption{Tx Code Online Adaptation}\label{alg_tx_adp_code}

\KwIn{Baseline model parameters $(\boldsymbol{\phi}, \boldsymbol{\theta}, \boldsymbol{\psi})$ trained on training set, the data sample to be transmitted $\boldsymbol{x}$, the CSI vector $\boldsymbol{h}$.}

\KwOut{The updated deep JSCC codeword $\boldsymbol{s}^{\star}$.}

\SetKwBlock{BeginRL}{procedure~\textnormal{\textsc{OnlineUpdateLatents}$(\boldsymbol{\phi}, \boldsymbol{\theta}, \boldsymbol{\psi}, \boldsymbol{x}, \boldsymbol{h}, \boldsymbol{y})$}}{}

\BeginRL{
Initialize latent representation: $\boldsymbol{y}^{(0)} \leftarrow \boldsymbol{y}$\;

\For{$t = 1,2,\dots,Y_{\max}$}
{
    Forward pass: ${{\boldsymbol{y}^{(t-1)}}}
\xrightarrow{{{f_e}( \cdot ;{\boldsymbol{\phi}}_f)}} {{\boldsymbol{s}}}
\xrightarrow{{{W}( \cdot|\boldsymbol{h})}} {{\boldsymbol{\hat s}}}
\xrightarrow{{{f_d}( \cdot ;{\boldsymbol{\theta}}_f)}} {{\boldsymbol{\hat y}}}
\xrightarrow{{{g_s}( \cdot ;{\boldsymbol{\theta}}_g)}} {{\boldsymbol{\hat x}}}$ with the latent prior ${{\boldsymbol{y}}^{(t-1)}} \xrightarrow{{{h_a}( \cdot ;{\boldsymbol{\phi}}_h)}} {{\boldsymbol{z}}}
\xrightarrow{{{h_s}( \cdot ;{\boldsymbol{\theta}}_h)}} \left\{ {{\boldsymbol{\mu}},{\boldsymbol{\sigma}}} \right\}$\;

    Compute loss $\mathcal{L}_{\text{R-D}}$ according to \eqref{eq_instant_loss_func}\;

    Update $\boldsymbol{y}^{(t)}$ using gradients $\nabla_{\boldsymbol{y}} \mathcal{L}_{\text{R-D}}$ as \eqref{eq_Tx_code_adp_iterative_a}\;

}

\Return Updated latent representation $\boldsymbol{y}^{\star} \leftarrow \boldsymbol{y}^{(Y_{\max})}$.
}

\SetKwBlock{BeginRC}{procedure~\textnormal{\textsc{OnlineUpdateCodewords}$(\boldsymbol{\phi}, \boldsymbol{\theta}, \boldsymbol{\psi}, \boldsymbol{x}, \boldsymbol{h}, \boldsymbol{s})$}}{}

\BeginRC{
Initialize deep JSCC codeword: $\boldsymbol{s}^{(0)} \leftarrow \boldsymbol{s}$\;

\For{$t = 1,2,\dots,S_{\max}$}
{
    Forward pass: ${{\boldsymbol{s}^{(t-1)}}}
\xrightarrow{{{W}( \cdot|\boldsymbol{h})}} {{\boldsymbol{\hat s}}}
\xrightarrow{{{f_d}( \cdot ;{\boldsymbol{\theta}}_f)}} {{\boldsymbol{\hat y}}}
\xrightarrow{{{g_s}( \cdot ;{\boldsymbol{\theta}}_g)}} {{\boldsymbol{\hat x}}}$\;

    Compute loss $\mathcal{L}_{\text{R-D}}$ according to \eqref{eq_instant_loss_func}\;

    Update $\boldsymbol{s}^{(t)}$ using gradients $\nabla_{\boldsymbol{s}} \mathcal{L}_{\text{R-D}}$ as \eqref{eq_Tx_code_adp_iterative_b}\;

}

\Return Updated deep JSCC codeword $\boldsymbol{s}^{\star} \leftarrow \boldsymbol{s}^{(S_{\max})}$.
}

\end{algorithm}

\subsection{Transceiver Adaptation}

The above transmitter adaptation method is appealing since no additional information needs to be added to the wireless transmitted signal, and nothing changes on the receiver. However, performance gains are relatively limited since the deep JSCC decoder $f_d$ and the nonlinear synthesis transform $g_s$ cannot be adapted. In this subsection, we present a method for transceiver \emph{full-model adaptation}, which tailors the entire NTSCC model to a specific domain. Unlike previous methods, our adaptive NTSCC with the full-model adaptation involves both \emph{data stream} and \emph{model stream} as the wireless transmitted signal. The model stream is utilized to inform the receiving end to update the model parameters of $f_d$ and $g_s$. \textit{A noteworthy point is that the transceiver full-model adaptation method is particularly suited for domain adaptation, as the model stream bandwidth cost associated with instance adaptation cannot be effectively amortized and results in inefficient computational complexity and high model stream bandwidth cost.} 
	
In this manner, the adapted model can be applied to other unseen samples within the same domain. Also, the full-model online learning process only takes place at the transmitter, as we assume the transmitter has a local copy of the pre-trained decoder models. This adaptation paradigm yields two kinds of tradeoffs:
\begin{enumerate}
	\item \emph{R-D-M trade-off:} a tripartite trade-off among the averaged data stream bandwidth cost ($R$), model stream bandwidth cost per domain ($M$), and distortion ($D$) terms, formulating the end-to-end R-D-M loss. Consider a specific source domain including $N$ samples, the model stream is transmitted only once, resulting in the actually averaged channel bandwidth cost of $k = R + M/N$, such that the averaged CBR is $\rho = k/m = (R + M/N)/m$.

	\item \emph{Performance-complexity trade-off:} the more constrained the domain adaptation, the greater potential gains from adaptation. However, a more restrictive domain necessitates multiple updating processes at the transmitter and multiple transmission model updates, leading to increased encoding complexity and delay.
\end{enumerate}

Transceiver full-model adaptation online updates a set of global baseline model parameters $(\boldsymbol{\phi}, \boldsymbol{\theta}, \boldsymbol{\psi})$ on a single source data instance $\boldsymbol{x}$. In practice, similar to that in the transmitter adaptation, we also fix the entropy model $h_a$ and $h_s$ to reduce the adaptation complexity. This results in the updated parameters $(\boldsymbol{\phi}_g^{\star}, \boldsymbol{\phi}_f^{\star}, \boldsymbol{\theta}_g^{\star}, \boldsymbol{\theta}_f^{\star})$, of which only $\boldsymbol{\theta}_g^{\star}$ and $\boldsymbol{\theta}_f^{\star}$ are transmitted over the wireless channel as the model stream.

Intuitively, a larger model steam bandwidth cost $M$ offers greater flexibility during model updating, potentially reducing the amortization gap for improved overall R-D performance. However, the actual performance, accounting for the model stream transmission cost, still requires further evaluation. As a result, developing an efficient model stream transmission method along is both challenging and crucial. Inspired by the residual coding idea, we transmit only the changes relative to the baseline model $\boldsymbol{\delta}_g = \boldsymbol{\theta}_g^{\star} - \boldsymbol{\theta}_g$ and $\boldsymbol{\delta}_f = \boldsymbol{\theta}_f^{\star} - \boldsymbol{\theta}_f$ in practice. To encode the model updates $\boldsymbol{\delta} = (\boldsymbol{\delta}_g, \boldsymbol{\delta}_f)$, we need to build a model prior $p_{\boldsymbol{\delta}}(\boldsymbol{\delta})$ to quantify the model rate. Accordingly, the model rate is derived with $-\log{p_{\boldsymbol{\delta}}(\boldsymbol{\delta})}$. Adding this term to the R-D loss function in \eqref{eq_instant_loss_func}, we obtain the full-model adaptive NTSCC objective:
\begin{equation}\label{eq_fullmodel_instant_loss_func}
\begin{aligned}
  & \mathcal{L}_{\text{R-D-M}}(\boldsymbol{\phi}, \boldsymbol{\theta}, \boldsymbol{\psi}, \boldsymbol{\delta}, \boldsymbol{x}, \boldsymbol{h}) \\
  & = \mathcal{L}_{\text{R-D}}(\boldsymbol{\phi}, \boldsymbol{\theta}+\boldsymbol{\hat \delta}, \boldsymbol{\psi}, \boldsymbol{x}, \boldsymbol{h}) + \beta \big(-{\eta_\delta}\log{p_{\boldsymbol{\delta}}(\boldsymbol{\delta})}\big) \\
  & = \lambda \underbrace{\big( -{\eta_y} \log{p_{\boldsymbol{ y}|\boldsymbol{ z}}(\boldsymbol{ y}|\boldsymbol{ z})} - {\eta_z}{\log{p_{\boldsymbol{ z}}(\boldsymbol{\tilde z})}} \big)}_{\text{data stream bandwidth cost:~}R} +  \underbrace{d(\boldsymbol{x},\boldsymbol{\hat{x}})}_{\text{distortion:~}D} \\
  & + \beta \underbrace{\big(-{\eta_\delta}\log{p_{\boldsymbol{\delta}}(\boldsymbol{\delta})}\big)}_{\text{model stream bandwidth cost:~}M},
\end{aligned}
\end{equation}
where the scaling factor $\eta_\delta$ is tied with the capability of codec used to transmit $\boldsymbol{\delta}$, $\boldsymbol{\hat \delta}$ denotes the reconstructed $\boldsymbol{\delta}$ at the receiver end, and $\beta$ controls the trade-off between the standard R-D loss and the model stream bandwidth cost. Minimization of $\mathcal{L}_{\text{R-D-M}}$ in \eqref{eq_fullmodel_instant_loss_func} ensures any cost in the model stream contributes to the R-D performance improvement.

For the model prior $p_{\boldsymbol{\delta}}$, any probability distribution function can be selected, herein, we naturally define $p_{\boldsymbol{\delta}}$ as a factorized model, i.e., $p_{\boldsymbol{\delta}}(\boldsymbol{\delta}) = \prod\nolimits_i{p_{{\delta_i}}({\delta_i})}$, where every component is of the same parameter. Each $p_{{\delta_i}}({\delta_i})$ is consistently generated from zero-centered Gaussian with a shared variance $\sigma^2$ as
\begin{equation}\label{eq_Gaussian_model}
\begin{aligned}
  p_{\boldsymbol{\delta}}(\boldsymbol{\delta}) & = \prod_i \underbrace{\left( \mathcal{N}(\delta_i|0,{{\sigma}}^2) * ({\Delta}\cdot \mathcal{U}(-\frac{\Delta}{2},\frac{\Delta}{2})) \right)}_{p_{\delta_i}} ({\delta}_i)\\
  ~ & = \prod_i \int_{{\delta}_i - \frac{\Delta}{2}}^{{\delta}_i + \frac{\Delta}{2}} {\mathcal{N}(\delta_i^\prime|0,{{\sigma}}^2)}{\rm d}{\delta_i^\prime},
\end{aligned}
\end{equation}
where the uniform distribution convolution is utilized to relax the prior such that the estimated model rate $-\log{p_{\boldsymbol{\delta}} (\boldsymbol{\delta})}$ stays non-negative. It can directly interpolate the discrete probability values $p_{\delta_i}({\bar \delta}_i)$ at the quantized values ${\bar \delta}_i$ that will be used when $\boldsymbol{\delta}$ is transmitted over the digital link with entropy coding and channel coding, and $\Delta$ in \eqref{eq_Gaussian_model} indicates the quantization bin width. If $\boldsymbol{\delta}$ needs to be quantized for entropy coding, we define the quantization function as that in \cite{van2021overfitting} with $N$ width-$\Delta$ quantization bins, i.e.,
\begin{equation}\label{eq_quantization_delta}
\begin{aligned}
  & {\bar \delta}_i = Q_{\Delta}({\delta_i}) \\
  & = {\rm{clip}}\Big({\left\lfloor \frac{\delta_i}{\Delta} \right\rceil} \cdot \Delta, \min = -\frac{(N-1)\Delta}{2},  \max = \frac{(N-1)\Delta}{2} \Big).
\end{aligned}
\end{equation}
During model training, the gradient of $Q_{\Delta}$ can be approximated by the Straight-Through estimator (STE) \cite{bengio2013estimating}. Correspondingly, we also leverage the uniformly-noised proxy $\boldsymbol{\tilde \delta} = \boldsymbol{\delta} + \boldsymbol{o}$ with $o_i\sim\mathcal{U}(-\frac{\Delta}{2},\frac{\Delta}{2})$ to compute the model stream rate during online training. In this case, the model stream rate is evaluated by substituting $\boldsymbol{\tilde \delta}$ into the model prior $p_{\boldsymbol{\delta}}$ as $-\log p_{\boldsymbol{\delta}}(\boldsymbol{\tilde \delta})$ in the R-D-M loss \eqref{eq_fullmodel_instant_loss_func}.

\begin{figure}[t]
	\begin{center}
		\hspace{-.05in}
		\subfigure[GMM density]{
			\includegraphics[scale=0.36]{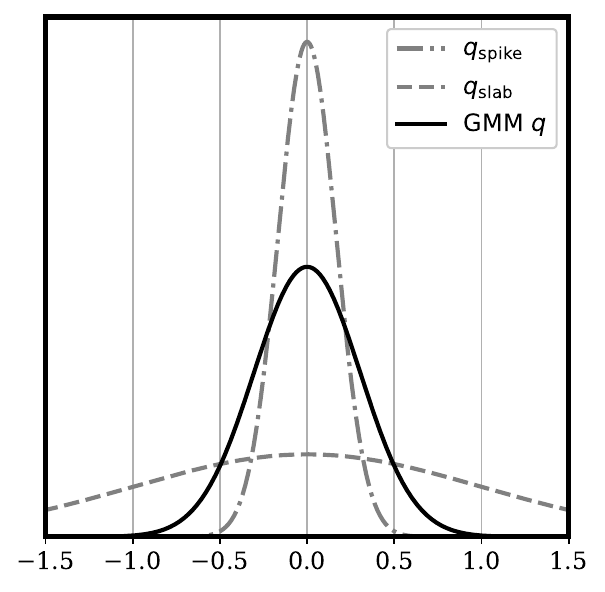}
		}
		\hspace{-.2in}
		\quad
		\subfigure[model prior]{
			\includegraphics[scale=0.36]{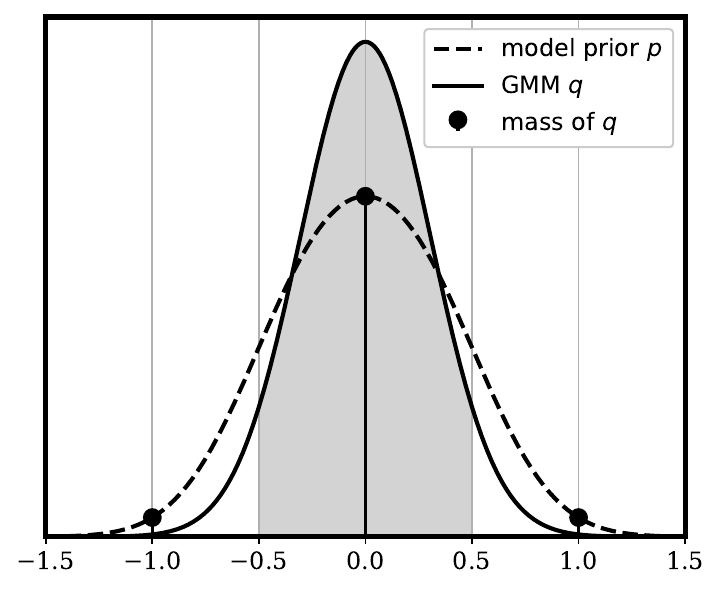}
		}
		
		\caption{A visual example of our GMM density function. The density $q_{\text{slab}}$ is generated with $\sigma = 1$, and the quantization bin width is set to $\Delta = 1$ such that the standard deviation $q_{\text{spike}}$ is $\frac{1}{6}$. The height of the spiky is $\alpha = 5$. Our model prior $p_{{\delta}_i}$ provides a continuous function, each value $p_{{\delta}_i}({{\bar \delta}_i})$ equals to the mass of GMM density $q_{\delta_i}(\delta_i)$ in the quantization bin $\Delta$ centered at ${{\bar \delta}_i}$.}
		\label{Fig_prob}
	\end{center}
	
	\vspace{0em}
\end{figure}

Note that most model updates $\delta_i$ (or the quantized ${\bar \delta}_i$) center around zero, the zero-centered Gaussian model prior in \eqref{eq_Gaussian_model} indeed ensures the minimum model stream cost for all-zero update, i.e., $-\log p_{\boldsymbol{\delta}}(\boldsymbol{0})$. To reduce the model stream cost, we can further generalize the standard Gaussian model prior as a \emph{Gaussian mixed model (GMM)} prior, which stems from \cite{van2021overfitting}. In particular, we adopt the widely-used spike-and-slab prior proposed in \cite{rovckova2018spike}, the model prior $p_{\boldsymbol{\delta}}$ will be generated using a weighted sum of two Gaussian distribution -- a wide (slab) Gaussian and a narrow (spike) Gaussian:
\begin{equation}\label{eq_GMM}
  q_{\boldsymbol{\delta}}({\boldsymbol{\delta}}) = \prod_i q_{\delta_i}(\delta_i) = \prod_i \frac{q_{\text{slab}}({\delta_i})+ \alpha q_{\text{spike}}({\delta_i})}{1+\alpha}
\end{equation}
with
\begin{equation}
  q_{\text{slab}}({\delta_i}) = \mathcal{N}({\delta_i} | {0}, {\sigma^2}),~q_{\text{spike}}({\delta_i}) = \mathcal{N}({\delta_i} | {0}, {(\frac{\Delta}{6})^2}),
\end{equation}
where $\alpha \ge 0 $ is a hyperparameter to determine the height of the spiky Gaussian with respect to the wider slab. Given $q_{\boldsymbol{\delta}}$, the model prior $p_{\boldsymbol{\delta}}$ is derived as
\begin{equation}\label{eq_GMM_model}
\begin{aligned}
  p_{\boldsymbol{\delta}}(\boldsymbol{\delta}) & = \prod_i \underbrace{\left( q_{\delta_i}(\delta_i) * ({\Delta}\cdot \mathcal{U}(-\frac{\Delta}{2},\frac{\Delta}{2})) \right)}_{p_{\delta_i}} ({\delta}_i)\\
  ~ & = \prod_i \int_{{\delta}_i - \frac{\Delta}{2}}^{{\delta}_i + \frac{\Delta}{2}} {q_{\delta_i}(\delta_i^\prime)}{\rm d}{\delta_i^\prime}.
\end{aligned}
\end{equation}
The discrete model prior $p_{{\delta}_i}({{\bar \delta}_i})$ is the pushforward of $p_{{\delta}_i}$ by substituting ${{\bar \delta}_i}$ into \eqref{eq_GMM_model}, which equals to the mass of GMM density $q_{\delta_i}(\delta_i)$ in the quantization bin $\Delta$ centered at ${{\bar \delta}_i}$. By setting the standard deviation of the spike as $\frac{\Delta}{6}$, we ensure the mass within the zero-centering bin $[-\frac{\Delta}{2}, \frac{\Delta}{2})$ cover $99.7\%$ of the total mass (``$3\sigma$-criterion''). A visual illustration of GMM function and model prior is shown in Fig. \ref{Fig_prob}.

The introduce of spike Gaussian distribution enforces sparsity on $f_d$ and $g_s$ model updates. A high spike weight (large $\alpha$) results in almost negligible bandwidth cost for transmitting the zero-update model stream. By this means, the full-model adaptation procedure can learn to make a binary decision: a parameter is worth updating by using higher bandwidth cost, or it is not updated by transmitting the negligible zero-update sign. Note that the whole online learning process takes place in the transmitter since we assume the transmitter has a local copy of the decoder parameters, thus the above binary decision is also made inside the transmitter. When the entropy coding is applied on ${\bar \delta}_i$, zero-update corresponds to a sequence of negligible `spent' bits. This trick can efficiently reduce the unnecessary model stream bandwidth cost.

The iterative transmitter model parameter updating procedure is the same as \eqref{eq_Tx_model_adp_iterative} by replacing $\mathcal{L}_{\text{R-D}}$ with $\mathcal{L}_{\text{R-D-M}}$, and the receiver model parameter updating procedures are written as
\begin{subequations}\label{eq_Rx_model_adp_iterative}
\begin{equation}\label{eq_Rx_model_adp_iterative_a}
  \boldsymbol{\theta}_g^{(t)} = \boldsymbol{\theta}_g^{(t-1)} - \gamma \nabla_{\boldsymbol{\theta}_g} \mathcal{L}_{\text{R-D-M}}(\boldsymbol{\phi}, \boldsymbol{\theta}, \boldsymbol{\psi}, \boldsymbol{\delta}, \boldsymbol{x}, \boldsymbol{h}),
\end{equation}
\begin{equation}\label{eq_Rx_model_adp_iterative_b}
   \boldsymbol{\theta}_f^{(t)} = \boldsymbol{\theta}_f^{(t-1)} - \gamma \nabla_{\boldsymbol{\theta}_f} \mathcal{L}_{\text{R-D-M}}(\boldsymbol{\phi}, \boldsymbol{\theta}, \boldsymbol{\psi}, \boldsymbol{\delta}, \boldsymbol{x}, \boldsymbol{h}).
\end{equation}
\end{subequations}
Since the model stream bandwidth cost $M$ is much smaller than the data stream bandwidth cost $R$, the transmission of model updates $\boldsymbol{\delta}$ default to a classical digital communication link, which will be quantized, entropy coded, channel coded, and modulated as a digital symbol sequence $\boldsymbol{m}$ passing over the wireless channel $W$. This transmission process is assumed to be reliable. Thus, the received $\boldsymbol{\hat \delta}$ is the quantized version $\boldsymbol{\bar \delta}$, and the model stream rate should be computed by substituting the uniformly-noised proxy $\boldsymbol{\tilde \delta}$ into the model prior $p_{\boldsymbol{\delta}}$, i.e., $-\log p_{\boldsymbol{\delta}}(\boldsymbol{\tilde \delta})$, in the R-D-M loss function $\mathcal{L}_{\text{R-D-M}}$. The encoding and decoding procedure of the transceiver full-model online learned NTSCC system is shown in Fig. \ref{Fig_full_model_system}. The whole online adaptation procedure is defined formally in Algorithm \ref{alg_txrx_adp_model}.

\begin{algorithm}[t]
	\setlength{\abovecaptionskip}{0cm}
	\setlength{\belowcaptionskip}{-0cm}
	
	\caption{TxRx Full-Model Online Adaptation}\label{alg_txrx_adp_model}
	
	\KwIn{Baseline model parameters $(\boldsymbol{\phi}, \boldsymbol{\theta}, \boldsymbol{\psi})$ trained on training set, model parameter quantizer $Q_{\Delta}$, model prior $p_{\boldsymbol{\delta}}$, the source dataset to be overfitted $\mathcal{X}$, the CSI vector $\boldsymbol{h}$.}
	
	\KwOut{The updated encoder models $\boldsymbol{\phi}_g^{\star}$ and $\boldsymbol{\phi}_f^{\star}$ and quantized decoder model updates $\boldsymbol{\bar \delta}$.}
	
	\SetKwBlock{BeginR}{procedure~\textnormal{\textsc{OnlineUpdateModels}$(\boldsymbol{\phi}, \boldsymbol{\theta}, \boldsymbol{\psi}, \boldsymbol{x}, \boldsymbol{h})$}}{}
	
	\BeginR{
		Initialize model parameters: $\boldsymbol{\phi}_g^{(0)} \leftarrow \boldsymbol{\phi}_g$, $\boldsymbol{\phi}_f^{(0)} \leftarrow \boldsymbol{\phi}_f$, $\boldsymbol{\theta}_g^{(0)} \leftarrow \boldsymbol{\theta}_g$, $\boldsymbol{\theta}_f^{(0)} \leftarrow \boldsymbol{\theta}_f$\;
		
		\For{$t = 1,2,\dots,T_{\max}$}
		{
			Quantize transmittable model parameters: $(\boldsymbol{\bar \theta}_g^{(t-1)},\boldsymbol{\bar \theta}_f^{(t-1)}) \leftarrow \boldsymbol{\bar \delta} + (\boldsymbol{\theta}_g,\boldsymbol{\theta}_f)$, with $\boldsymbol{\bar \delta} = Q_{\Delta}(\boldsymbol{\delta})$ and $\boldsymbol{\delta} = (\boldsymbol{\theta}_g^{(t-1)},\boldsymbol{\theta}_f^{(t-1)}) - (\boldsymbol{\theta}_g,\boldsymbol{\theta}_f)$\;
			
			Draw a batch of data samples ${\boldsymbol{x}}$ from the source domain $\mathcal{X}$;
			
			\Comment{specifically used for domain adaptation}
			
			Forward pass: ${{\boldsymbol{x}}} \xrightarrow{{{g_a}( \cdot ;{\boldsymbol{\phi}}_g^{(t-1)})}} {{\boldsymbol{y}}}
			\xrightarrow{{{f_e}( \cdot ;{\boldsymbol{\phi}}_f^{(t-1)})}} {{\boldsymbol{s}}}
			\xrightarrow{{{W}( \cdot|\boldsymbol{h})}} {{\boldsymbol{\hat s}}}
			\xrightarrow{{{f_d}( \cdot ;{\boldsymbol{\bar \theta}}_f^{(t-1)})}} {{\boldsymbol{\hat y}}}
			\xrightarrow{{{g_s}( \cdot ;{\boldsymbol{\bar \theta}}_g^{(t-1)})}} {{\boldsymbol{\hat x}}}$ with the latent prior ${{\boldsymbol{y}}} \xrightarrow{{{h_a}( \cdot ;{\boldsymbol{\phi}}_h)}} {{\boldsymbol{z}}}
			\xrightarrow{{{h_s}( \cdot ;{\boldsymbol{\theta}}_h)}} \left\{ {{\boldsymbol{\mu}},{\boldsymbol{\sigma}}} \right\}$\;
			
			Compute loss $\mathcal{L}_{\text{R-D-M}}$ according to \eqref{eq_fullmodel_instant_loss_func} by setting $\boldsymbol{\hat \delta} = \boldsymbol{\bar \delta}$ and the model rate proxy $-\log p_{\boldsymbol{\delta}}(\boldsymbol{\tilde \delta})$\;
			
			Backpropagate using STE for $Q_{\Delta}$, and update $(\boldsymbol{\phi}_g^{(t)}$, $\boldsymbol{\phi}_f^{(t)}, \boldsymbol{\theta}_g^{(t)}$, $\boldsymbol{\theta}_f^{(t)})$ using gradients $(\nabla_{\boldsymbol{\phi}_g} \mathcal{L}_{\text{R-D-M}}, \nabla_{\boldsymbol{\phi}_f} \mathcal{L}_{\text{R-D-M}}, \nabla_{\boldsymbol{\theta}_g} \mathcal{L}_{\text{R-D-M}}, \nabla_{\boldsymbol{\theta}_f} \mathcal{L}_{\text{R-D-M}})$ as \eqref{eq_Tx_model_adp_iterative} and \eqref{eq_Rx_model_adp_iterative}\;
			
		}
		
		\Return Updated models $\boldsymbol{\phi}_g^{\star} \leftarrow \boldsymbol{\phi}_g^{(T_{\max})}$ and  $\boldsymbol{\phi}_f^{\star} \leftarrow \boldsymbol{\phi}_f^{(T_{\max})}$, and quantized model updates $\boldsymbol{\bar \delta}$.
	}
	
\end{algorithm}

\begin{figure*}[t]
	\setlength{\abovecaptionskip}{0.cm}
	\setlength{\belowcaptionskip}{-0.cm}
	\centering{\includegraphics[scale=0.26]{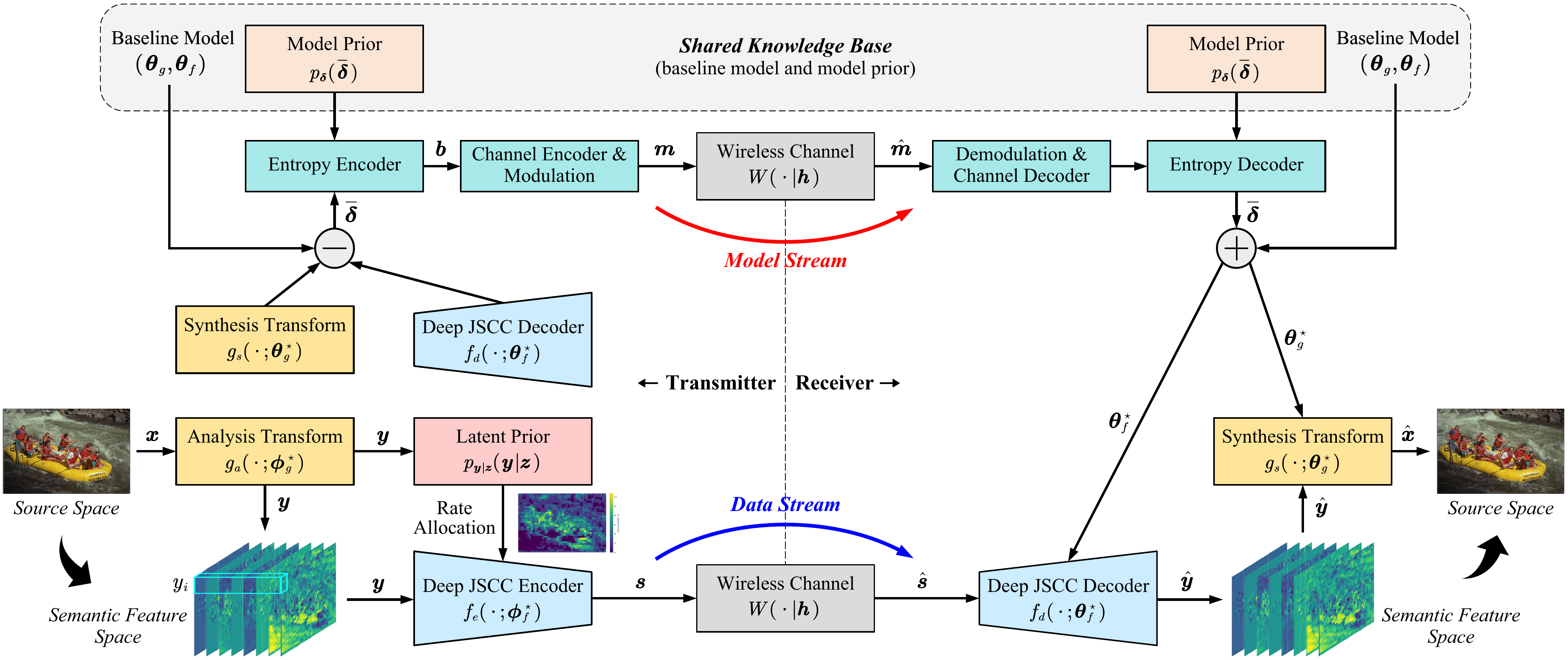}}
	\caption{Visualization of encoding and decoding of our full-model online learned NTSCC system.}\label{Fig_full_model_system}
\end{figure*}

\section{Overfitting the Channel}\label{section_channel_overfit}

\begin{figure*}[t]
	\centering
	\includegraphics[scale=0.95]{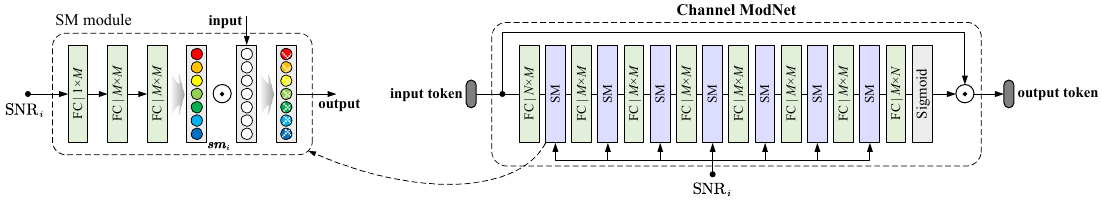}
	\caption{Channel-dependent deep JSCC codec used for overfitting the CSI instance in our adaptive NTSCC system. FC denotes fully-connected network, and the following parameter marks the ``$\text{input dimension} \times \text{output dimension}$''. ``$\odot$'' denotes the element-wise product.}\label{Fig_channel_overfit}
	\vspace{0em}
\end{figure*}

In this section, we discuss how to adapt to the instant CSI $\boldsymbol{h} \in {\mathcal{H}}^\prime$ such that the channel domain adapted model in the previous Section \ref{section_source_overfit} can be further upgraded to the instance adapted model, i.e.,
\begin{subequations}\label{eq_channel_overfit_goal}
\begin{equation}
    {\boldsymbol{\Sigma}}_{(\boldsymbol{x},\mathcal{H}^\prime)} \Rightarrow {\boldsymbol{\Sigma}}_{(\boldsymbol{x},\boldsymbol{h})},
\end{equation}
\begin{equation}
   {\boldsymbol{\Sigma}}_{(\mathcal{X}^\prime,\mathcal{H}^\prime)} \Rightarrow {\boldsymbol{\Sigma}}_{(\mathcal{X}^\prime,\boldsymbol{h})}.
\end{equation}
\end{subequations}
In can be observed that the value of instant $\boldsymbol{h}$ is a key factor affecting the performance of deep JSCC codec. Under different channel states, different resource allocation strategies should be adopted to implicitly adjust source coding rate and channel coding rate inside the JSCC codec. The vanilla NTSCC model \cite{dai2022nonlinear} is trained under the objective \eqref{eq_instant_loss_func} using different signal-to-noise ratio (SNR) indicating different channel states. Recently, an attention module was introduced into convolutional neural networks (CNN) to make deep JSCC SNR-adaptive using a single model \cite{xu2021wireless}. However, a simple average SNR cannot fully capture all the characteristics of the channel state $\boldsymbol{h}$. To tackle this, a CA-JSCC scheme was proposed in \cite{wu2022channel} that employs dual-attention mechanism to achieve adaptation over multipath OFDM channels. Even though these methods have proposed novel adaptation strategies, they are mostly based on traditional convolutional auto-encoders. In this paper, we investigate a new channel-dependent mechanism specifically tailored for vision Transformers (ViTs) adopted in NTSCC, enabling the whole system automatically to adapt to $\boldsymbol{h}$ without relying on gradient descent.

Our idea is introducing a \emph{plug-in} module to modulate the output of ViT-based deep JSCC encoder in NTSCC as shown in Fig. \ref{Fig_channel_overfit}. The proposed ``Channel ModNet'' is a plug-in module inserted as the last layer of deep JSCC encoder or the first layer of deep JSCC decoder. For different channel states $\boldsymbol{h}$, we obtain different neural-syntax to generate a more specific deep JSCC codec functions $f_e$ and $f_d$.

In particular, the architecture of the proposed Channel ModNet for instant channel state $\boldsymbol{h}$ adaptation is depicted in Fig. \ref{Fig_channel_overfit}. ModNet consists of $8$ FC layers separated by $7$ SNR modulation (SM) modules. SM is a three-layered FC network with input being the channel SNR that corresponds to $y_i$ (denoted as ${\rm{SNR}}_i$). As stated before, the receiver can obtain the instant CSI $\boldsymbol{h}$ via channel estimation, thus the ModNet in the deep JSCC decoder $f_d$ can obtain the explicit ${\rm{SNR}}_i$ for each $y_i$. Herein, ${\rm{SNR}}_i$ is computed by averaging the channel symbol SNRs along the ${\bar k}_i$-dimensional sequence $s_i$. To be aligned with practical systems, we assume the CQI available at the transmitter via a feedback link, which represents an averaged SNR along all transmitted symbols $\boldsymbol{s}$. Thus, the SM module input is ${\rm{SNR}}_i = {\rm{\overline{SNR}}}$ at the transmitter. The SM module transforms the input ${\rm{SNR}}_i$ into an $c$-dimensional tensor $\boldsymbol{sm}_i$ as depicted in Fig. \ref{Fig_channel_overfit}. In this manner, the arbitrary target modulator can be realized by assigning a corresponding SNR value. And the CSI $\boldsymbol{h}$ is therefore associated with the $c$-dimensional tensor $\boldsymbol{sm}_i$ in each SM module. Then, the input $c$-dimensional feature will be fused with $\boldsymbol{sm}_i$ in the element-wise product.

Multiple SM modules are cascaded sequentially in a coarse-to-fine manner, as shown in Fig. \ref{Fig_channel_overfit}. The previously modulated features are fed into subsequent SM modules. In our design, the channel state modulation is comprehensively considered in a token-wise attention fashion. In this way, our channel-dependent modulation method can notify the JSCC codec what the current channel state is. The proposed channel-dependent method can be applied to arbitrary wireless channel model, and the ModNet trained under a specific channel domain $\mathcal{H}^\prime$ can provide better channel adaptation performance for the instances $\boldsymbol{h} \in \mathcal{H}^\prime$.

\section{Experimental Results}\label{section_performance}

In this section, we first outline the experimental configurations and evaluation protocol, followed by the presentation of source/channel overfitting evaluation results for the online learned NTSCC. Also, we carry out a comprehensive ablation study and provide analytical discussions to highlight the key advantages and clarify some limitations of this work.

\subsection{Experimental Setup}

\subsubsection{Datasets}

Regarding the wireless video transmission, we adopt the Video Streaming Dataset 4K (VSD4K) dataset \cite{liu2021overfitting} for evaluation. The VSD4K dataset is composed of several 4K 30fps videos, including six popular video categories: \texttt{game}, \texttt{vlog}, \texttt{interview}, \texttt{sport}, \texttt{dance}, and \texttt{city}. The first three categories are mainly single-scene but have multiple points of view, while the latter contains various scenes and large-scale motion. In this paper, we consider the I-frame wireless transmission problem, i.e., the test set consists a group of I-frame images from a single video of a specific category, which is a subproblem of video transmission. We select a 45s-length representative video from each category, subsample each video by the first of four frames to create a dataset of I-frames, and generate low-resolution videos of 480P, 720P, and 1080P using bicubic interpolation. Unless otherwise specified, the following experiments will be carried out on 480P video sources.

\subsubsection{Network Implementation}

Note that the proposed source and channel overfitting method are model agnostic that can be transplanted to any neural network based end-to-end transmission system, we employ NTSCC \cite{dai2022nonlinear} as a representative architecture for online updating in this paper. As stated before, we do not transmit the side information $\boldsymbol{\bar{z}}$ to simplify the implementation of NTSCC. We use the DIV2K \cite{agustsson2017ntire} image set to train our baseline model, which contains 800 natural images of 2K resolutions on average. The baseline model training procedure includes 6,000 epochs using Adam optimizer \cite{kingma2014adam} with the learning rate $1 \times 10^{-4}$. During model training, the images are randomly cropped to $256  \times 256$ patches to form a batch of $8$ patches. In all experiments, we train four baseline models with $\lambda = 256, 64, 16, 4$ respectively under the additive white Gaussian noise (AWGN) channel, where the scaling factor $\eta_y$ is set to $0.2$ at $\text{SNR}=10$dB and $0.4$ at $\text{SNR}=0$dB. The mean squared error (MSE) is used as the distortion measurement to qualify the end-to-end image transmission performance. For subjective comparison, we further consider perceptual quality optimization to better align with the human vision in semantic communication system. The training details of NTSCC (Perceptual) are the same as that in \cite{dai2022nonlinear}.

\subsubsection{Transmitter Adaptation Details}

The Tx-adaptation schemes comprise two main approaches, namely the online transmitter model updating and the transmitter code updating. The former can be applied either to every I-frame (i.e., instance adaptation) or to every video sequence (i.e., domain adaptation) within the VSD4K dataset. The latter is applicable only in the instance adaptive mode. Specifically, for the instance adaptation, the transmitter model adaptation scheme updates ($g_a$, $f_e$) for $T_{\max}=100$ steps per I-frame with the learning rate $\gamma = 1 \times 10^{-4}$. Regarding transmitter code adaptation, we first update $\boldsymbol{y}$ for $Y_{\max}=50$ steps and then update $\boldsymbol{s}$ for another $S_{\max}=50$ steps, where the learning rate is set to $\gamma = 1 \times 10^{-3}$ in both optimization procedures. It is possible to update with more steps using the same or lower learning rate, but our experiments have shown that negligible performance gain can be obtained.

\subsubsection{Transceiver Adaptation Details}

The TxRx-adaptation scheme builds upon the aforementioned pre-trained 4 baseline models, we online update the parameters $(\boldsymbol{\phi}_g, \boldsymbol{\phi}_f, \boldsymbol{\theta}_g, \boldsymbol{\theta}_f)$ for the R-D-M loss $\mathcal{L}_{\text{R-D-M}}$ over each video sequence. To ensure the average channel bandwidth cost of updated models close to the baseline, the R-D-M trade-off hyperparameter tuples $(\lambda, \beta)$ is set to $(256,4)$, $(32,4)$, $(4,1)$, and $(2,1)$, which correspond to the baseline models learned with $\lambda = 256, 64, 16, 4$, respectively. For the decoder updating process, which will be transmitted to the receiver as the model stream, we set the parameter quantization bin width as $\Delta = 0.005$ and utilize the GMM prior, where $\sigma$ in $q_{\text{slab}}$ is set to $0.05$, and the spike weight $\alpha = 1000$. The number of updating steps is set to $T_{\max} = 10000$, with the learning rate of $\gamma = 1 \times 10^{-5}$ for $g_a$ and $f_e$, and $\gamma = 1 \times 10^{-4}$ for $g_s$ and $f_d$. We select the best updated model based on the R-D-M loss for model inference.

\subsubsection{Comparison Schemes}

\begin{figure*}[t]
	\centering
	\subfigure[]{
		\includegraphics[scale=0.3]{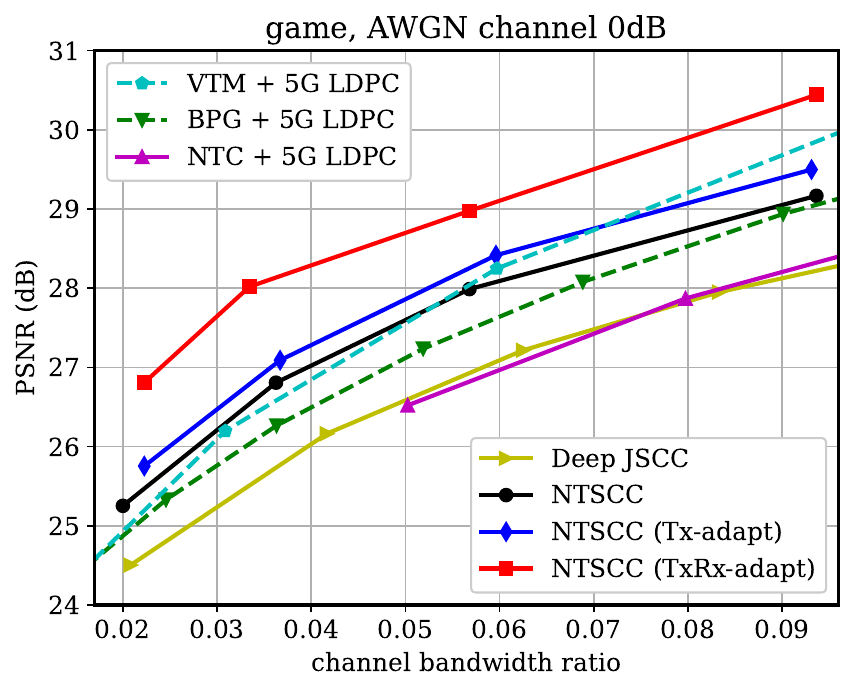}
	}
	\hspace{-.15in}
	\subfigure[]{
		\includegraphics[scale=0.3]{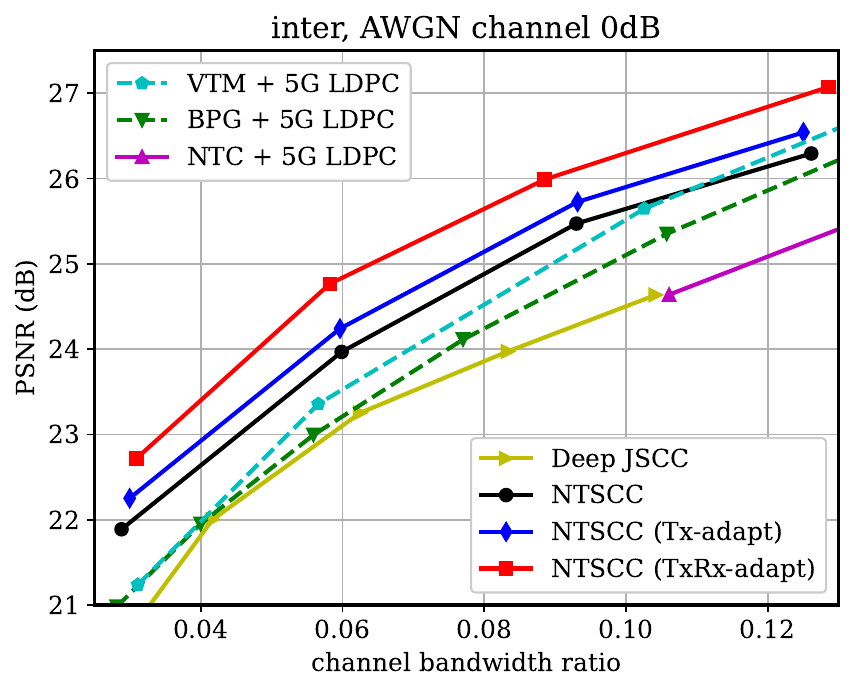}
	}
	\hspace{-.15in}
	\subfigure[]{
		\includegraphics[scale=0.3]{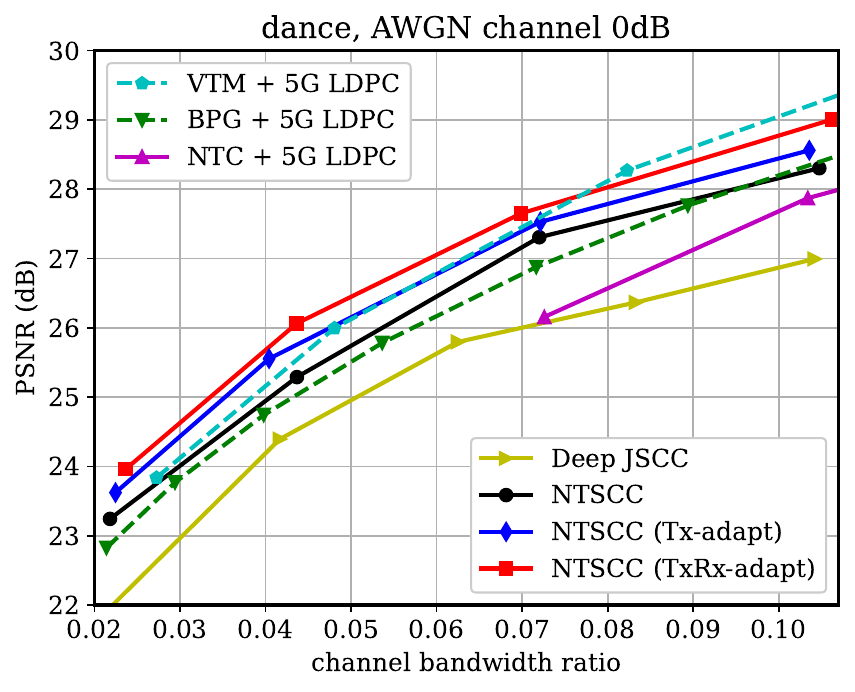}
	}
	\hspace{-.15in}
	\subfigure[]{
		\includegraphics[scale=0.3]{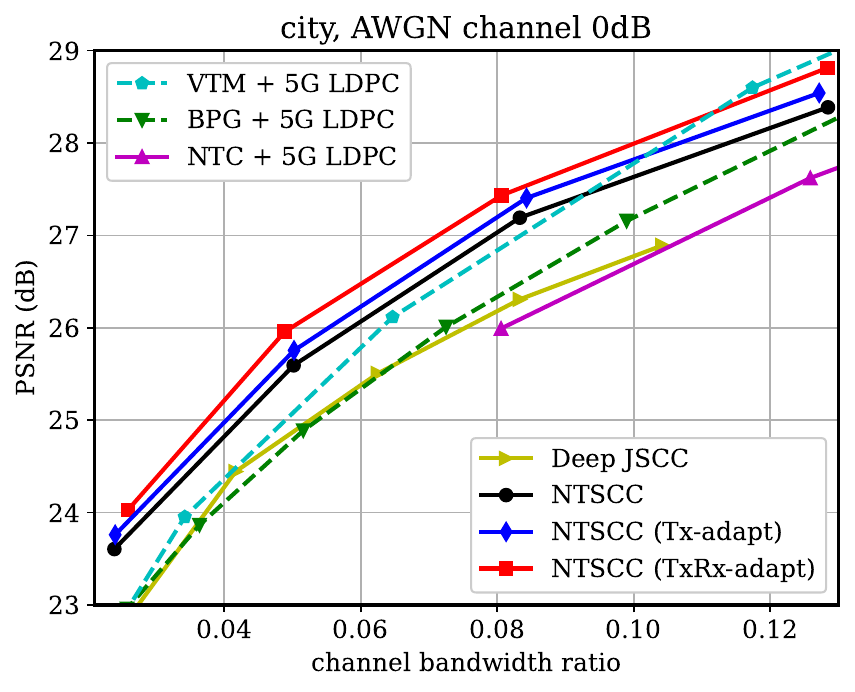}
	}
	\subfigure[]{
		\includegraphics[scale=0.3]{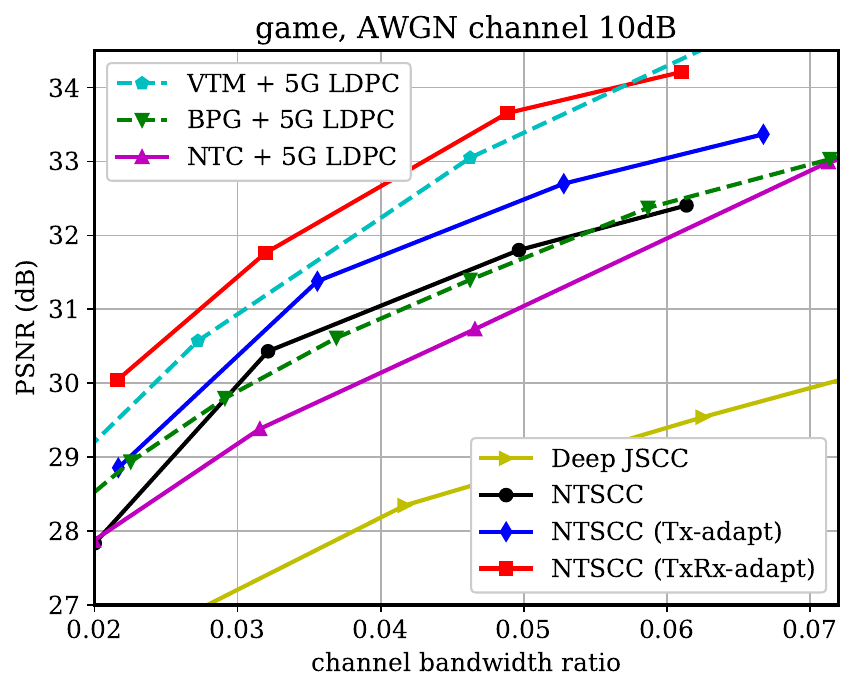}
	}
	\hspace{-.15in}
	\subfigure[]{
		\includegraphics[scale=0.3]{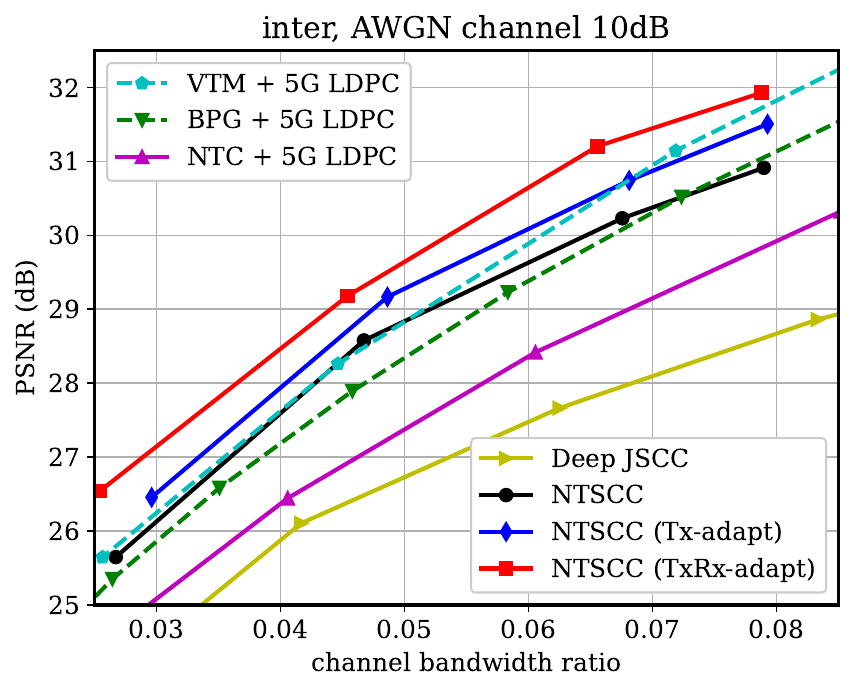}
	}
	\hspace{-.15in}
	\subfigure[]{
		\includegraphics[scale=0.3]{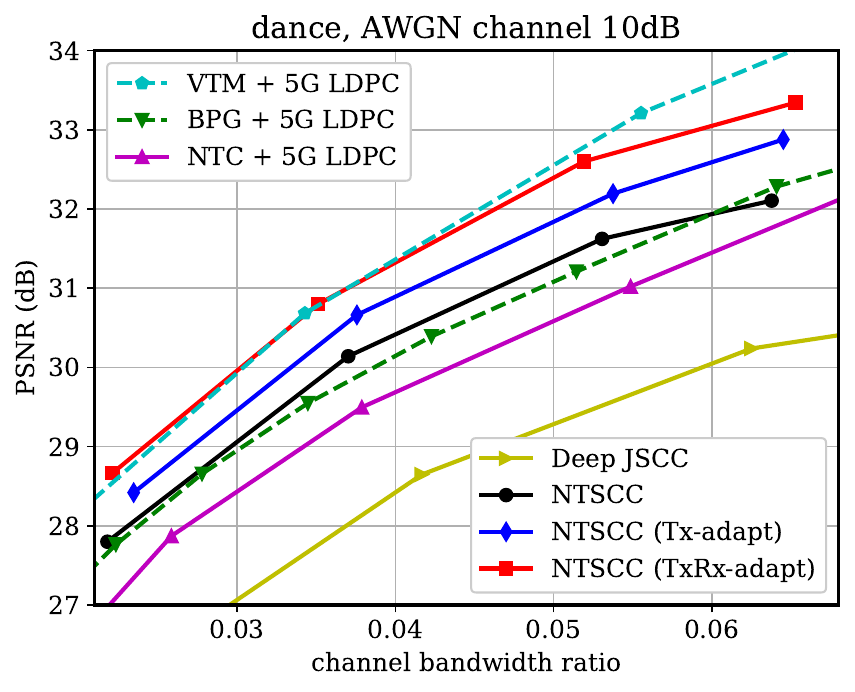}
	}
	\hspace{-.15in}
	\subfigure[]{
		\includegraphics[scale=0.3]{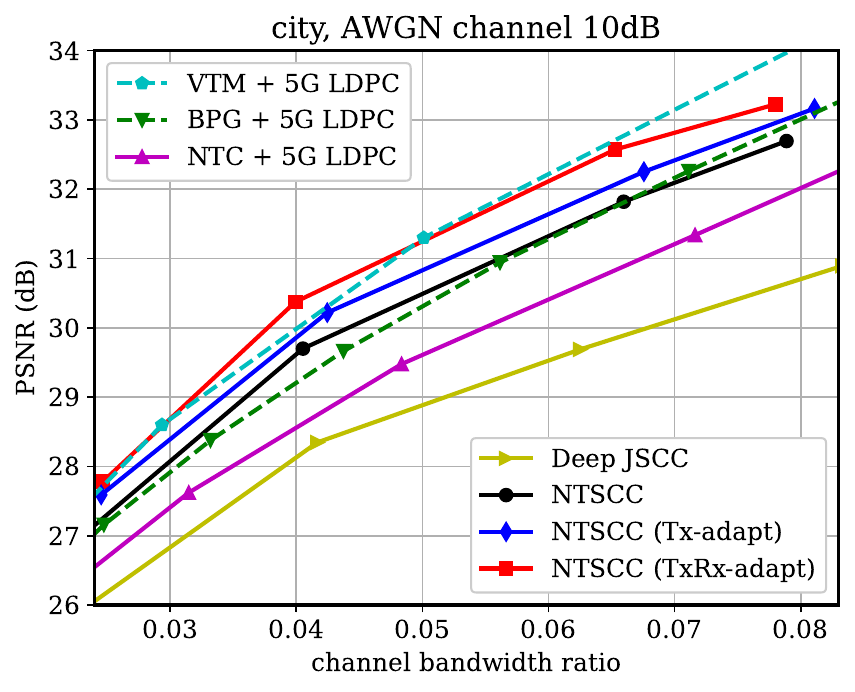}
	}
	\caption{PSNR performance versus the average channel bandwidth ratio (CBR) $\rho$ over the AWGN channel at $\text{SNR} = 0$dB (bad channel condition, subfigures (a)--(d)) and $\text{SNR} = 10$dB (good channel condition, subfigures (e)--(h)). In this figure, our adaptive NTSCC systems are online updated toward each video sequence, which is sequence domain adaptation.}
	\label{Fig_rdres}
\end{figure*}

Apart from the comparison with baseline NTSCC system, we also compare our online learned NTSCC with SOTA engineered wireless video transmission systems. Specifically, we compare our approach with HEVC (H.265) \cite{sullivan2012overview} and VVC (H.266) \cite{bross2021developments} for video compression combined with 5G LDPC code \cite{richardson2018design} for channel coding. We use ``$+$'' operator to combine source coding and channel coding schemes for brevity. For VVC and HEVC, we use the official test model VTM-12.2 with intra profile and BPG software to encode I-frames in the YUV444 mode, and calculate the PSNR in RGB. Additionally, the NTC source compression scheme proposed in \cite{balle2018} combined with 5G LDPC codes are also included in comparison. We adopt the released NTC models based on mean and scale hyper-prior implemented by CompressAI \cite{begaint2020compressai}. To ensure a fair comparison, we crop all I-frames to multiples of 64 to avoid padding for neural codecs. The above simulations are implemented on the top of Sionna \cite{hoydis2022sionna}, which is a open-source library for link-level simulations of digital communication systems. Apart from these methods, we also compare our online learned NTSCC with the emerging neural deep JSCC transmission scheme \cite{DJSCCF}. The channel bandwidth cost counts up all the transmitting streams over the wireless channel, including both data and model in our full-model adaptation mode.

\subsection{Objective Performance Analysis}

Fig. \ref{Fig_rdres}(a)--(h) show the averaged R-D performance under AWGN channels with $\text{SNR} = 0$dB (poor channel condition) and $\text{SNR} = 10$dB (good channel condition), respectively. In Fig. \ref{Fig_rdres}, the legend ``NTSCC (TxRx-adapt)'' denotes NTSCC with our transceiver full-model domain adaptation, and ``NTSCC (Tx-adapt)'' denotes our transmitter model instance adaptation. For the classical separation-based schemes, according to the adaptive modulation coding (AMC) mechanism, we need to traverse given combinations of LDPC coded modulation schemes to identify the highest-efficiency scheme under the reliable transmission constraint (block error rate $\le 10^{-5}$). Accordingly, we adopt a $1/3$ rate $(2048, 6144)$ LDPC code with 4QAM at $\text{SNR} = 0$dB, and a $2/3$ rate $(4096, 6144)$ LDPC code with 16-ary quadrature amplitude modulation (16QAM) at $\text{SNR} = 10$dB. Results in Fig. \ref{Fig_rdres} indicate that by using our proposed online overfitting mechanisms, we achieve considerable gains compared to the baseline NTSCC, and the transceiver full-model adaptation results in higher performance gain than updating the transmitter only. Moreover, compared to the best performed engineered transmission system ``VTM + 5G LDPC'', our NTSCC (TxRx-adapt) shows comparable or even better performance. As one typical scheme towards adaptive semantic communications, our online learned NTSCC shows the potential becoming the emerging SOTA scheme for end-to-end wireless data transmission. The existing well-known improvements of semantic communication are compared with ``BPG + 5G LDPC'' \cite{dai2022nonlinear} or on the tiny-scale image dataset \cite{DJSCC, DJSCCF}, our performance gain is more meaningful as the first end-to-end data transmission scheme overpassing the SOTA coded transmission system (VTM + 5G LDPC) on high-resolution image/video datasets.

\begin{figure*}[t]
	\centering
	\subfigure[CBR percentage, 0dB]{
		\includegraphics[scale=0.23]{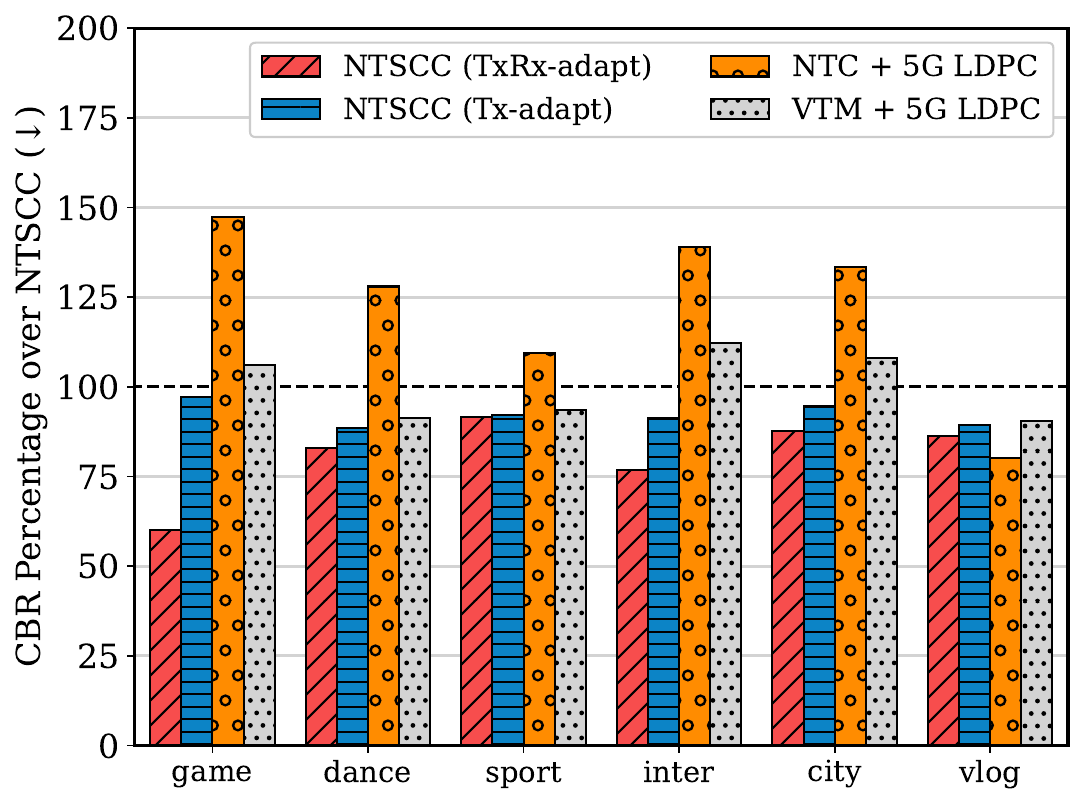}
	}
	\hspace{-.1in}
	\subfigure[CBR percentage, 10dB]{
		\includegraphics[scale=0.23]{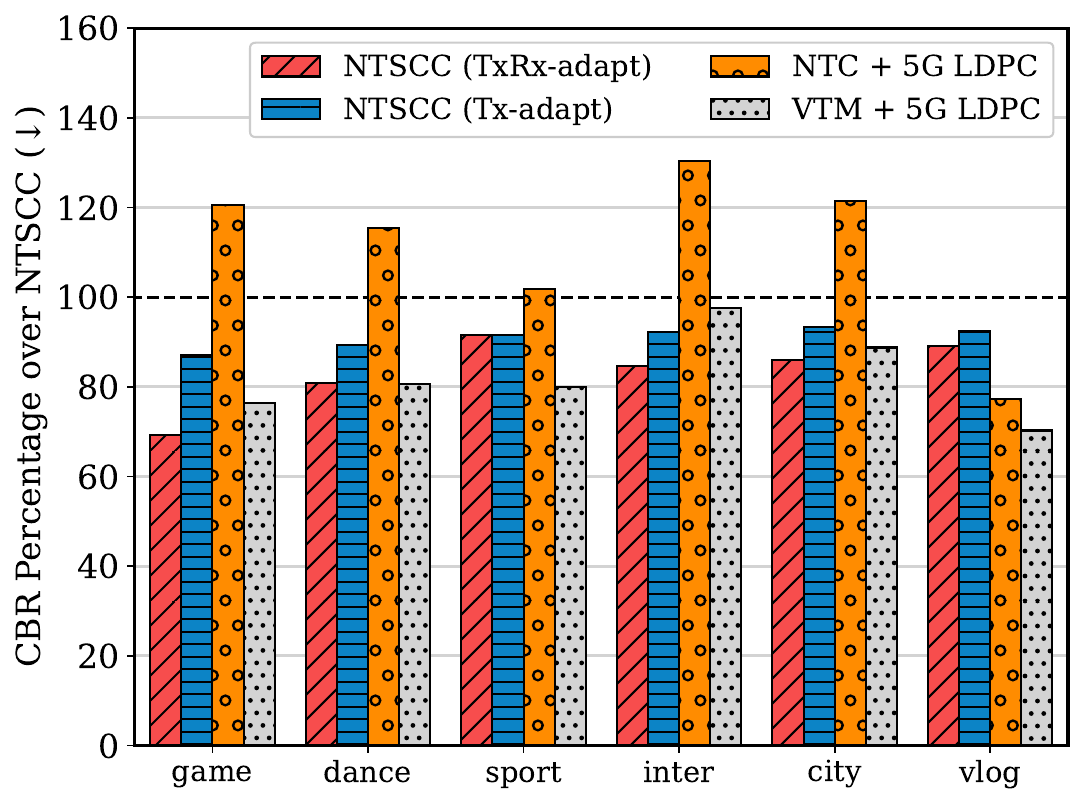}
	}
	\hspace{-.1in}
	\subfigure[PSNR gain, 0dB]{
		\includegraphics[scale=0.23]{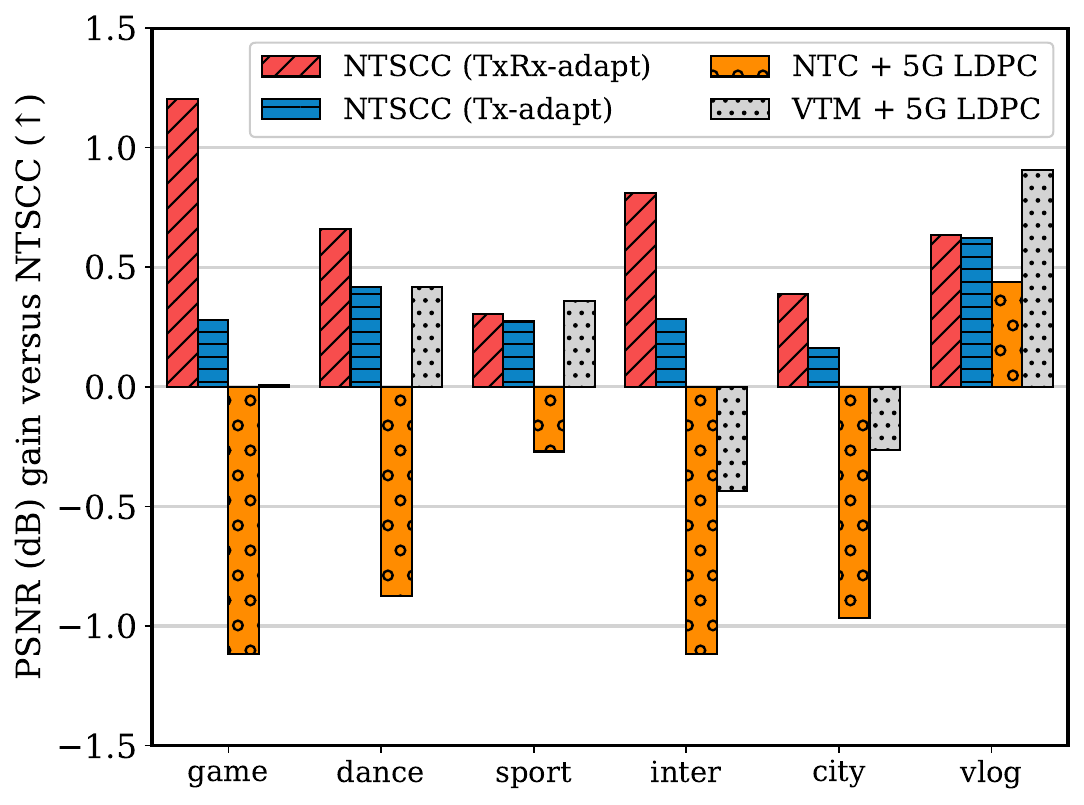}
	}
	\hspace{-.1in}
	\subfigure[PSNR gain, 10dB]{
		\includegraphics[scale=0.23]{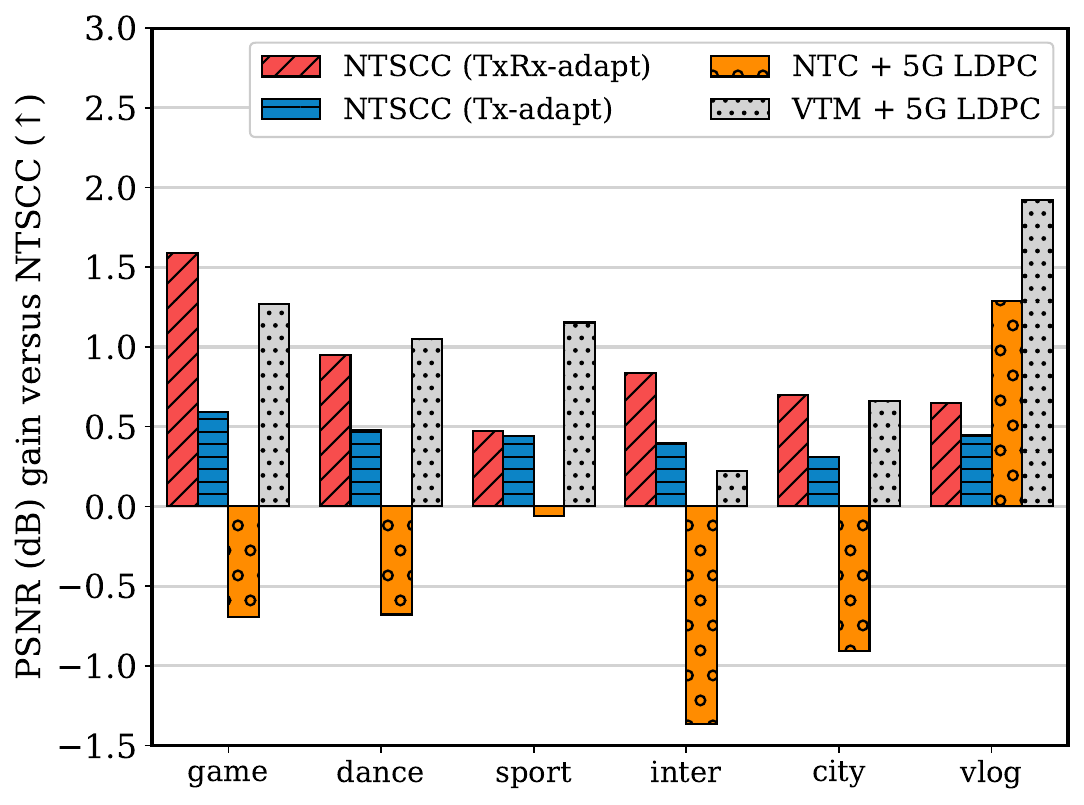}
	}
	\caption{Average CBR percentage and PSNR gains versus the baseline NTSCC semantic communication system over the AWGN channel at $\text{SNR} = 0$dB and $\text{SNR} = 10$dB, where ``$\downarrow$'' on the Y-axis label denotes ``lower is better'', and ``$\uparrow$'' denotes ``higher is better''.}
	\label{Fig_res_gains}
\end{figure*}

We further plot the PSNR gain or bandwidth saving of the six video sequences in Fig. \ref{Fig_res_gains}. Here, we employ the widely-used Bj{\o}ntegaard Delta (BD) rate reduction algorithm \cite{bjontegaard2001calculation} for relative performance evaluation. The baseline scheme is NTSCC trained on the DIV2K dataset. Fig. \ref{Fig_res_gains}(a) and \ref{Fig_res_gains}(b) show the relative average CBR percentage over the baseline at the same PSNR under channel SNR $0$dB and $10$dB, respectively. Fig. \ref{Fig_res_gains}(c) and \ref{Fig_res_gains}(d) show the average quality improvement in terms of PSNR over the baseline for each scheme. Specifically, NTSCC with transceiver adaptation (TxRx-adapt) can save up to $41\%$ bandwidth cost, while NTSCC with only transmitter adaptation (Tx-adapt) can save up to $13\%$. Among the six sequences in VSD4K, our NTSCC (TxRx-adapt) performs better than all other systems for most categories (4/6). However, it shows slightly worse performance on the \texttt{sport} and \texttt{vlog} sequences. We attribute this performance loss to the presence of multiple shots and diverse scenarios in these two sequences, which complicates the adaptation of the baseline NTSCC model to the entire video sequence. To improve performance, it is recommend to restrict the adaptation range to align with the obvious scene changes within a video sequence.

\begin{figure}
	\centering
	\includegraphics[scale=0.5]{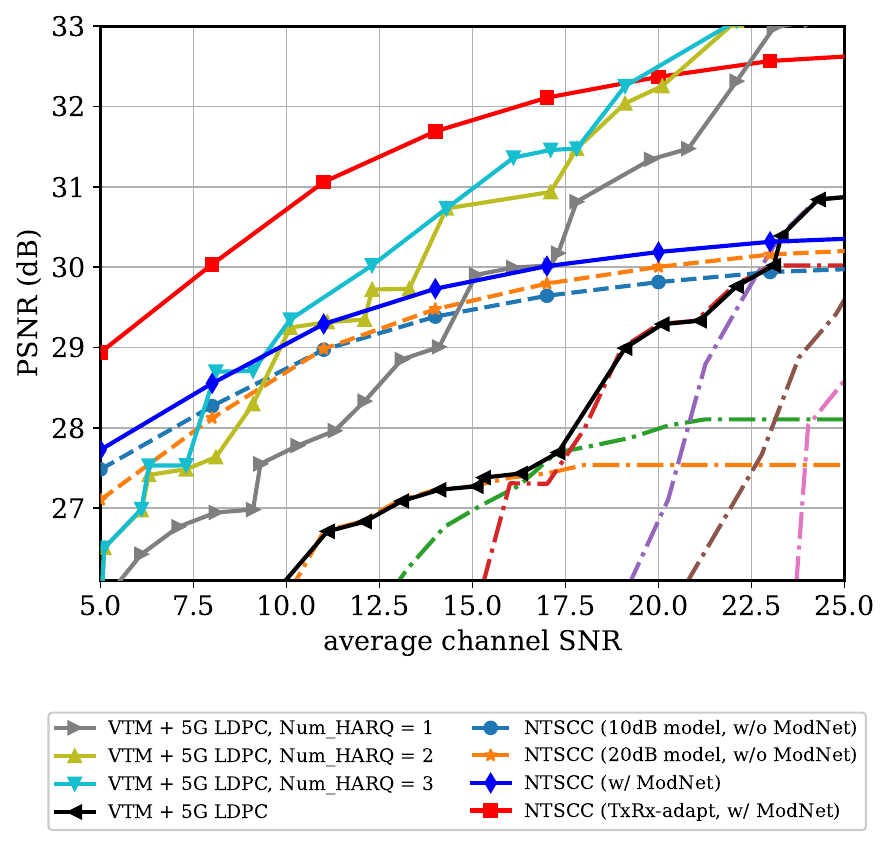}
	\caption{PSNR performance versus the change of average SNR over the COST2100 5.3GHz indoor channel under the CBR constraint $\rho \leq 0.03$. The two baseline NTSCC models are trained under this COST2100 fading channel with $\text{average SNR} = 10$dB and $\text{average SNR} = 20$dB, respectively.}
	\label{Fig_cost2100}
\end{figure}

\makeatletter
\renewcommand{\@thesubfigure}{\hskip\subfiglabelskip}

Next, we verifies the flexibility of our proposed plug-in Channel ModNet on adaptation to the instant channel state. We consider the widely-used COST2100 wireless fading channel model \cite{liu2012cost} for evaluation. CSI samples are collected in an indoor scenario at 5.3GHz bands. In this case, the transmitted symbols $\boldsymbol{s}$ will pass over a frequency selective channel on the OFDM grid. We transmit the I-frames of \texttt{game} video sequence over this channel.

Fig. \ref{Fig_cost2100} shows the channel adaptation results. It can be seen that NTSCC with ModNet (see the solid blue curve) performs better than the standard NTSCC model (without ModNet) trained at a given average SNR. It verifies the Channel ModNet module can make NTSCC model aware of the instant token-wise SNRs, which is more accurate than using a rough average SNR among all embeddings.

Moreover, performance can be further improved by utilizing full-model adaptation based on the NTSCC with ModNet, as illustrated  by the red line in Fig. \ref{Fig_cost2100}. This approach enables both source domain adaptation and channel instance adaptation. Additionally, we also present a comparison with VTM + 5G LDPC schemes using the same CSI samples collected from COST2100. The configurations of LDPC codes and QAM are consistent with those in \cite{DJSCCF}, and we take the envelope of all combinations of coded transmission schemes as the final performance (black line in Fig. \ref{Fig_cost2100}). Apparently, it cannot provide satisfactory quality using one-shot transmission due to the cliff-effect. To align better with practical communication system, we further adopt the hybrid automatic repeat request (HARQ) with chase combining (CC) \cite{ahmed2021hybrid} to enhance the system performance while our NTSCC schemes continues to use one-shot transmission. It can be seen that retransmissions bring considerable PSNR gains especially for the high SNR region, and the gain increases with the number of HARQ allowed. However, HARQ also introduces much higher transmission latency. As a comparison, the TxRx-adapt NTSCC (red line in Fig. \ref{Fig_cost2100}) can provide competitive performance with one-shot transmission.

\subsection{Visual Examples and User Study}

\begin{figure}[t]
	\centering
	\hspace{-.1in}
	\subfigure[]{
		\includegraphics[scale=0.28]{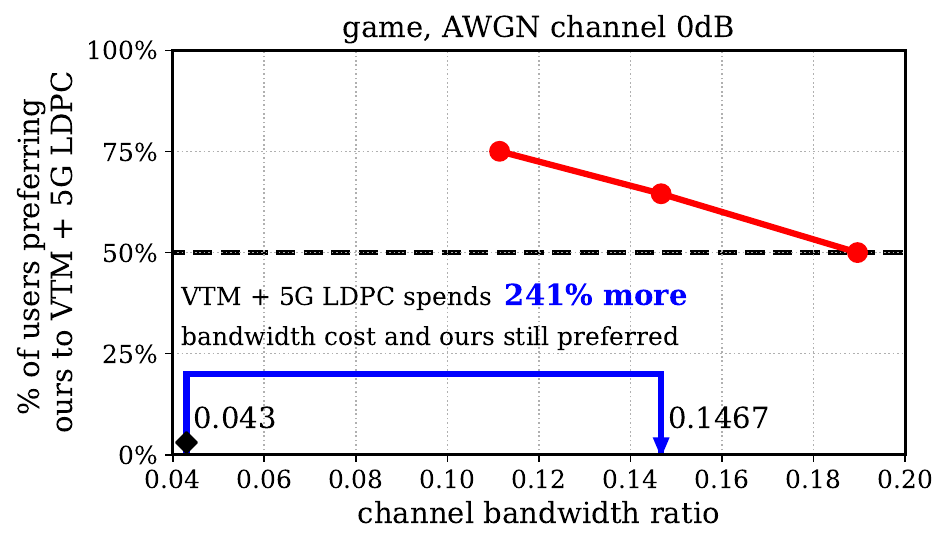}
	}
	\hspace{-.15in}
	\subfigure[]{
		\includegraphics[scale=0.28]{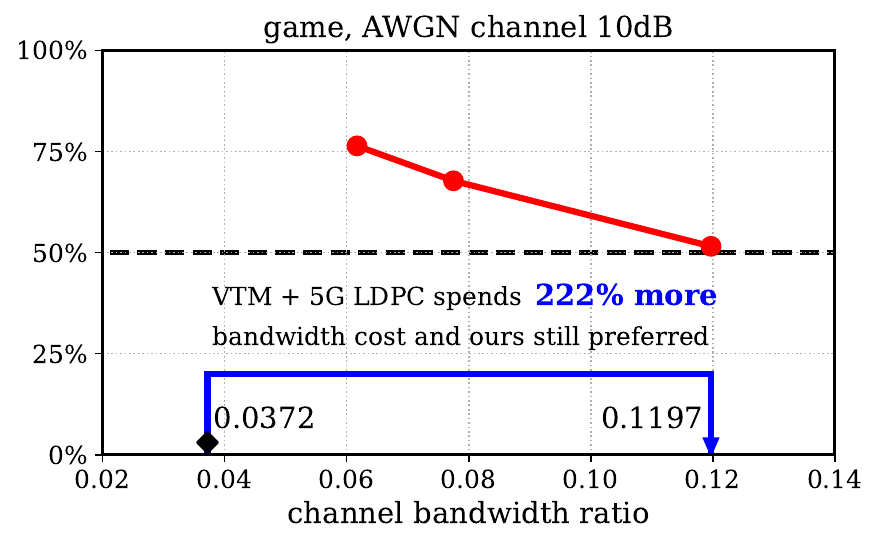}
	}
	
	\hspace{-.1in}
	\subfigure[]{
		\includegraphics[scale=0.28]{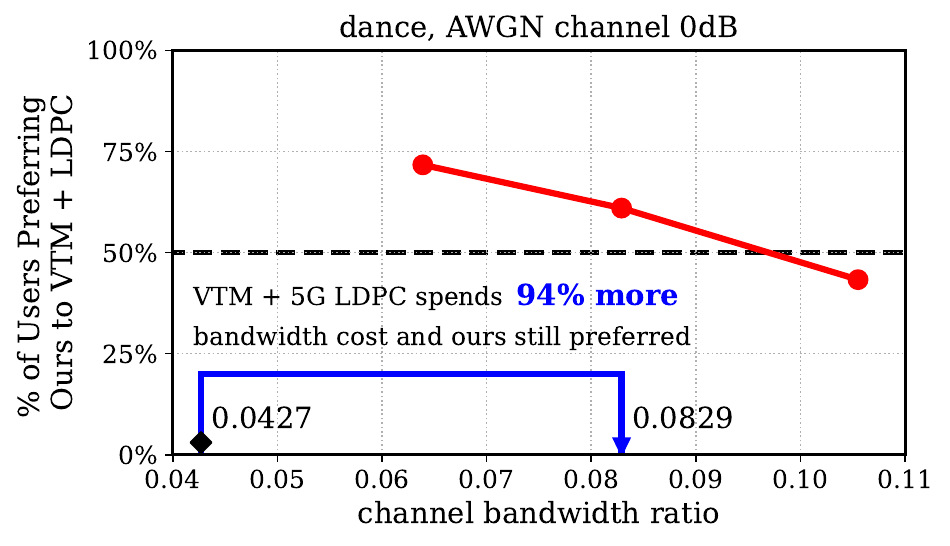}
	}
	\hspace{-.15in}
	\subfigure[]{
		\includegraphics[scale=0.28]{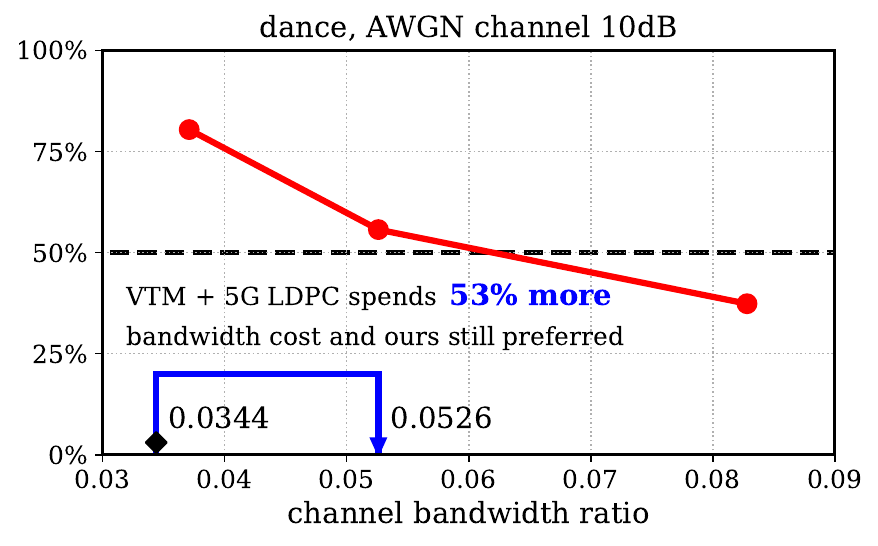}
	}
	
	\hspace{-.1in}
	\subfigure[]{
		\includegraphics[scale=0.28]{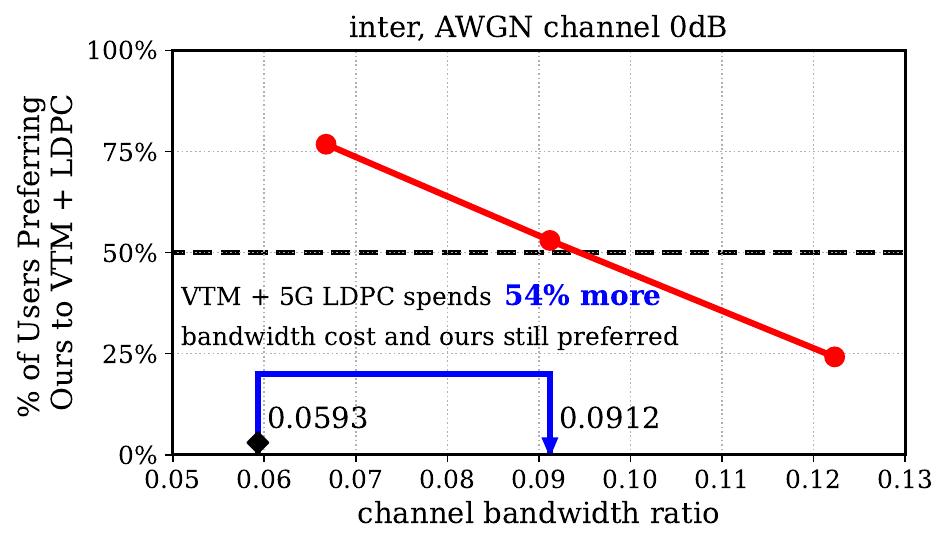}
	}
	\hspace{-.15in}
	\subfigure[]{
		\includegraphics[scale=0.28]{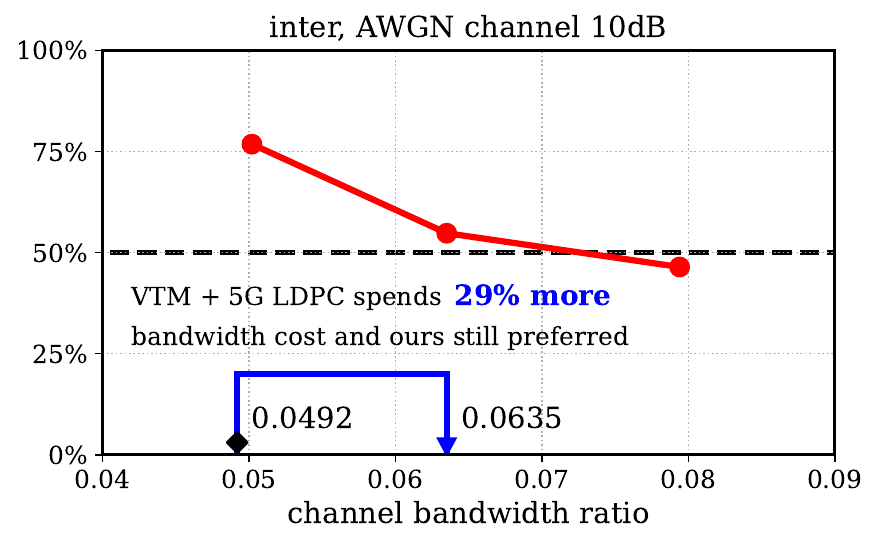}
	}
	
	\hspace{-.1in}
	\subfigure[]{
		\includegraphics[scale=0.28]{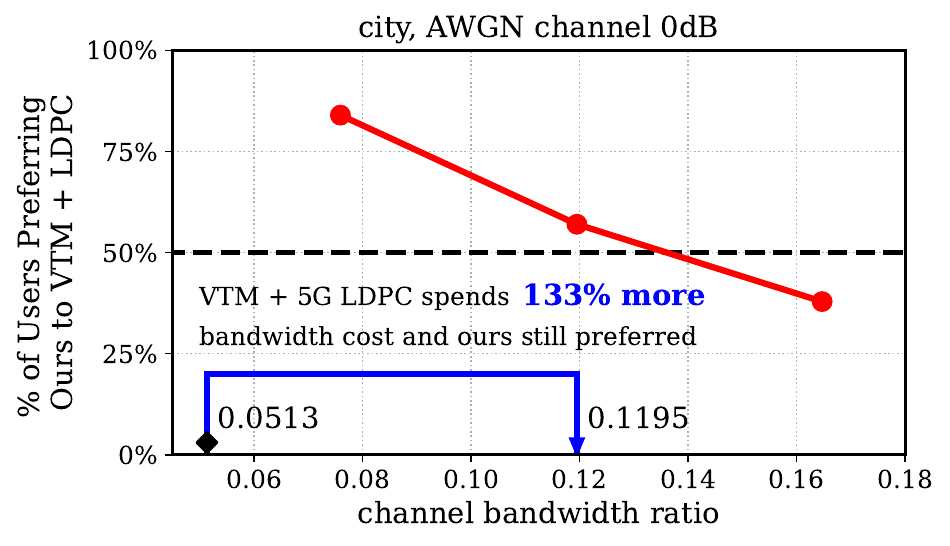}
	}
	\hspace{-.15in}
	\subfigure[]{
		\includegraphics[scale=0.28]{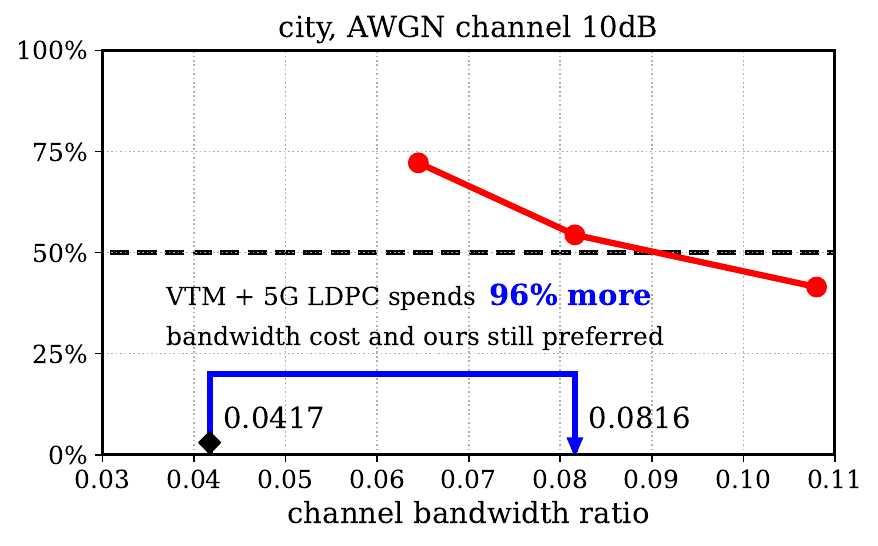}
	}
	\caption{User study results. As an anchor, the black diamond point denotes the CBR cost of our adaptive NTSCC (TxRx-adapt) semantic communication system optimized under perceptual loss. For each data point in the red line, its Y-axis shows what percentage of users prefer our scheme, and the X-axis is the CBR cost of VTM + 5G LDPC. The blue arrow highlights how much extra CBR cost VTM + 5G LDPC spends where still more than 50\% of participants prefer ours. The results are counted from 2953 ratings in total, with an average of 123 per data point.}
	\label{Fig_user_study}
\end{figure}

\begin{figure*}
	
	\begin{subtable}
		
		\centering
		\small
		\begin{tabular}{m{0.185\textwidth}m{0.165\textwidth}<{\centering}m{0.178\textwidth}<{\centering}m{0.17\textwidth}<{\centering}m{0.17\textwidth}<{\centering}}
			
		\centering  & \centering VTM + 5G LDPC  & \centering \textbf{Online learned NTSCC} \\ \footnotesize{\textbf{(TxRx-adapt, Perceptual)}}  & \centering Online learned NTSCC \\ \footnotesize{(TxRx-adapt, PSNR)} & \centering Online learned NTSCC \\ \footnotesize{(Tx-adapt, PSNR)} \tabularnewline
			
		\end{tabular}
	\end{subtable}
	\vspace{-.03in}
	\begin{center}
		\hspace{-.00in}
		\subfigure[$704\times 1280$] {\includegraphics[width=0.19\textwidth]{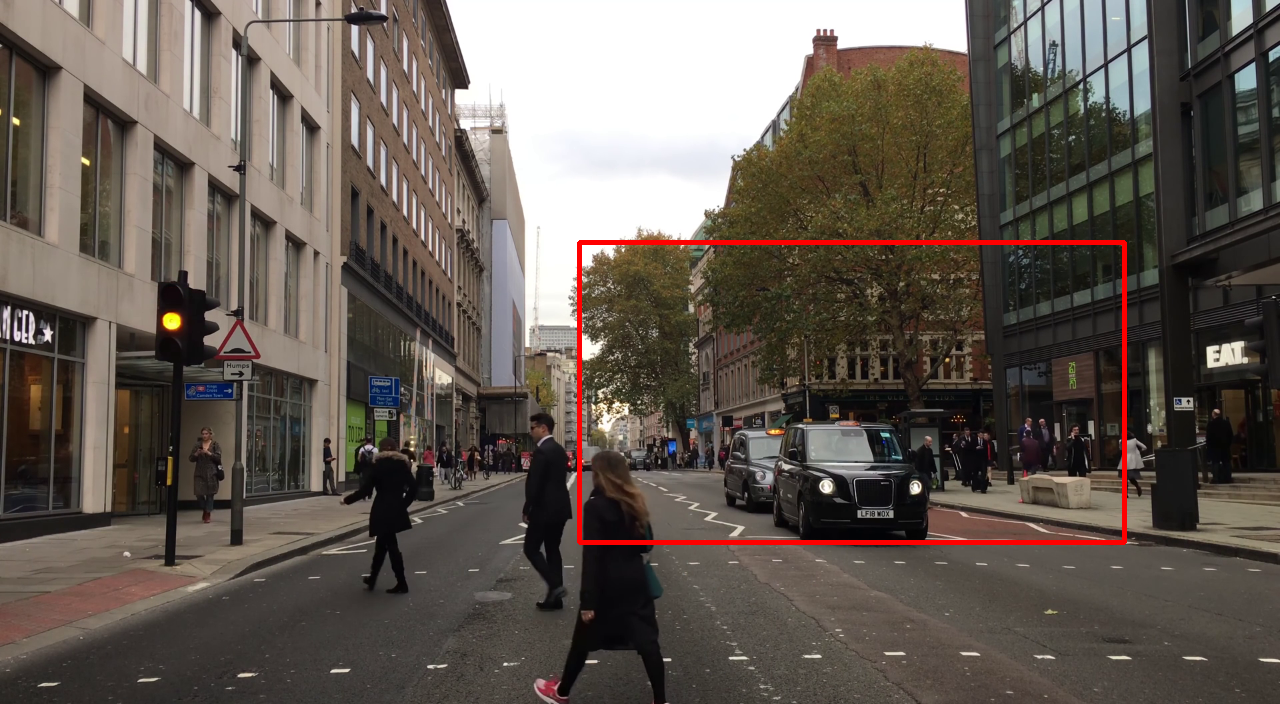}}
		\hspace{-.20in}
		\quad
		\subfigure[0.0381 (\textit{0\%}) / 28.38] {\includegraphics[width=0.19\textwidth]{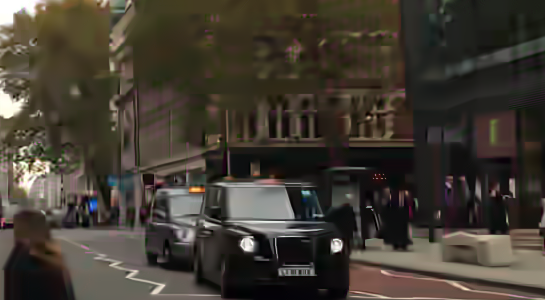}}
		\hspace{-.20in}
		\quad
		\subfigure[0.0327 (\textcolor{blue}{\textit{--14.2\%}}) / 28.29]{\includegraphics[width=0.19\textwidth]{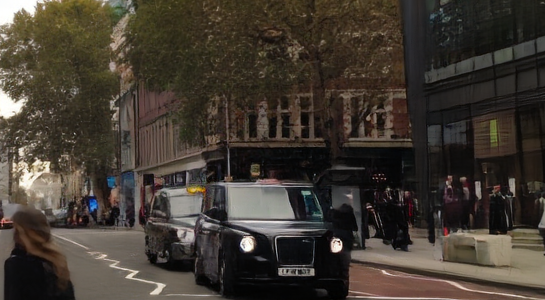}}
		\hspace{-.20in}
		\quad
		\subfigure[0.0330 (\textcolor{blue}{\textit{--13.4\%}}) / 28.78] {\includegraphics[width=0.19\textwidth]{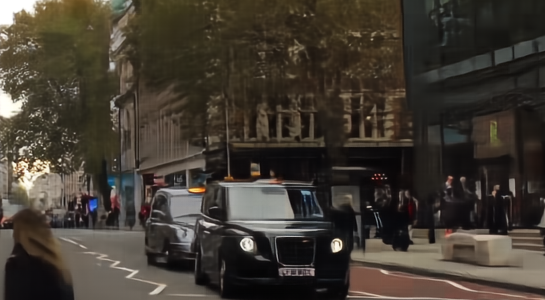}}
		\hspace{-.20in}
		\quad
		\subfigure[0.0355 (\textcolor{blue}{\textit{--6.8\%}}) / 28.49] {\includegraphics[width=0.19\textwidth]{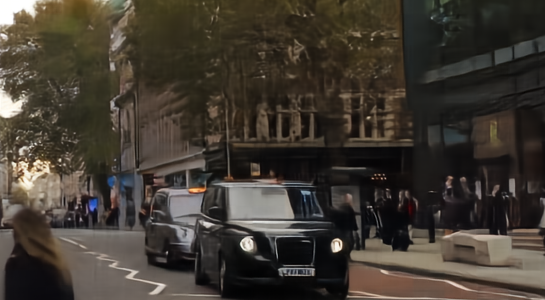}}
		\hspace{-.20in}
	\end{center}
	
	\begin{subtable}
		
		\centering
		\small
		
		\begin{tabular}{m{0.201\textwidth}m{0.15\textwidth}<{\centering}m{0.188\textwidth}<{\centering}m{0.157\textwidth}<{\centering}m{0.185\textwidth}<{\centering}}
			\centering Original & \centering BPG + 5G LDPC & \centering Deep JSCC & \centering NTC + 5G LDPC & \centering NTSCC\tabularnewline
			
		\end{tabular}
	\end{subtable}
	\vspace{-.16in}
	\begin{center}	
		\hspace{-.20in}
		\subfigure[$R$ / PSNR (dB)] {\includegraphics[width=0.19\textwidth]{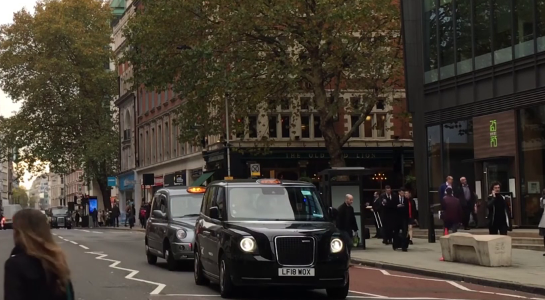}}
		\hspace{-.20in}
		\quad
		\subfigure[0.0358 (\textcolor{blue}{\textit{--6.0\%}}) / 27.65] {\includegraphics[width=0.19\textwidth]{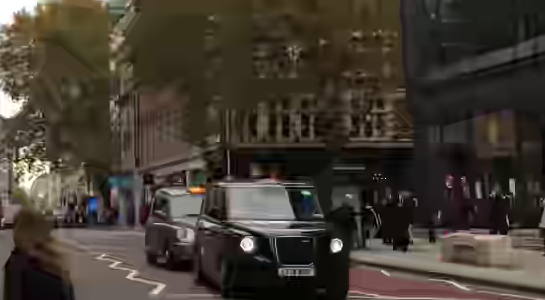}}
		\hspace{-.20in}
		\quad
		\subfigure[0.0414 (\textcolor{red}{\textit{+3.5\%}}) / 27.68]{\includegraphics[width=0.19\textwidth]{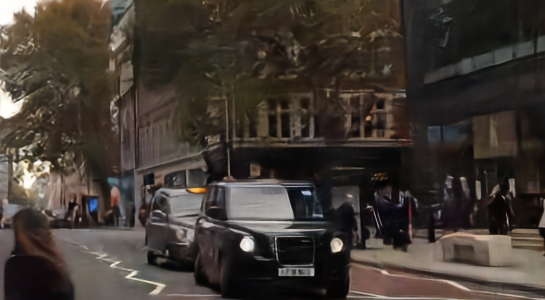}}
		\hspace{-.20in}
		\quad
		\subfigure[0.0420 (\textcolor{red}{\textit{+10.2\%}}) / 28.11] {\includegraphics[width=0.19\textwidth]{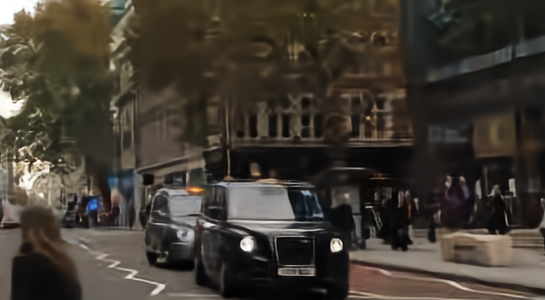}}
		\hspace{-.20in}
		\quad
		\subfigure[0.0348 (\textcolor{blue}{\textit{--8.7\%}}) / 28.13] {\includegraphics[width=0.19\textwidth]{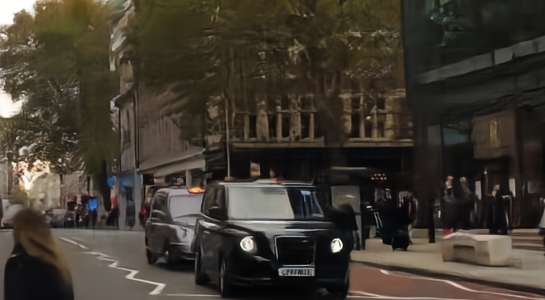}}
		\hspace{-.20in}
		
		\caption{Examples of visual comparison. The first column shows the original image and its cropped patch. The second to the fifth column show the reconstructed images by using different transmission schemes over the AWGN channel at $\text{SNR} = 0$dB, where the metrics in parentheses indicate the model training target loss function (PSNR or perceptual loss). Red number and blue number indicate the percentage of bandwidth cost increase and saving compared to the ``VTM + 5G LDPC'' scheme.}
		\label{Fig_visual}
	\end{center}
\end{figure*}

Since the quality metrics sometimes fail to account for many nuances of human perception, we therefore rely on human opinions collected in a thorough user study. We employ the ``two alternatives, forced choice'' (2AFC) test for quantitative evaluation. As a widely-used approach in the perceptual visual quality assessment \cite{HIFIC,zhang2018unreasonable,agustsson2019}, the user study interface shows human raters a group of three images in the same column, where the middle image is always the original image, and the other two are lossily generated from two different methods. Then, we require raters to choose which image (left or right) looks closer to the original. We adopt the NTSCC (TxRx-adapt) model learned under perceptual loss \cite{dai2022nonlinear} and compare it to VTM + 5G LDPC with close or more CBR cost. In particular, we randomly collect $10$ I-frame images for each video sequence transmitted over the AWGN channel at $\text{SNR} = 0$dB and $\text{SNR} = 10$dB. The entire rating process consists of several rounds in which all participants were asked to vote on each of the $10$ groups of images. If our NTSCC (TxRx-adapt) dominates, the CBR cost of VTM + 5G LDPC will be increased in the next round until it receives more votes. User study results are given in Fig. \ref{Fig_user_study}. As we can see, our source and channel overfitting approach enables NTSCC to greatly outperform the SOTA VTM + 5G LDPC transmission scheme, especially for \texttt{game} and \texttt{city} sequences. It verifies that the proposed online learned NTSCC system can better support the human perceptual vision demands in wireless data transmission, which is aligned with the target of semantic communications.

To intuitively demonstrate the effect of the proposed overfitting method, we further pick visible results on transmitting the \texttt{city} video sequence in Fig. \ref{Fig_visual}. From these results, we can observe that online learned NTSCC optimized under the perceptual loss achieves much better visual quality than other schemes with lower channel bandwidth cost.

\subsection{Ablation Study}

\subsubsection{R-D-M Tradeoff}

Table \ref{table_R-D-M_tradeoff} provides an ablation study concerning the impact of $\beta$ in domain adaptation, which investigates the tripartite trade-off among the data stream bandwidth cost ($R$), model stream bandwidth cost ($M$), and end-to-end distortion ($D$). Specifically, we demonstrate the distribution of total CBR cost using a fixed $\lambda=4$ in a 45s \texttt{game} sequence (including $N=372$ I-frames) over the AWGN channel at $\text{SNR} = 10$dB. Herein, the model stream cost $M$ is counted according to the channel state and the AMC mechanism, as that used for the traditional separation-based schemes in Fig. \ref{Fig_rdres}. When the LDPC coding rate and digital modulation order are determined, the number of transmitted symbols $M$ can be calculated according to the estimated model stream entropy in \eqref{eq_fullmodel_instant_loss_func}, \eqref{eq_GMM_model} and the selected AMC scheme. From this table, we observe that as the value of $\beta$ decreases, the R-D terms start to dominate the R-D-M trade-off, resulting in improved R-D performance compared to the baseline model (in the first row of Table \ref{table_R-D-M_tradeoff}) and an increased bandwidth cost for the model stream. since the decoder is updated only once to adapt to the \texttt{game} domain, the bandwidth cost of the model stream can be amortized over a substantial number of I-frames. As a result, the model rate becomes negligible, especially in the high $\beta$ region. An appropriate selection of $\beta$ in the transceiver adaptation can contribute to a significant improvement in reconstruction quality while also reducing the total bandwidth cost $k=R+\frac{M}{N}$.

\begin{table}[t]
	\renewcommand{\arraystretch}{1.3}
	\centering
	\small
	
	\caption{R-D-M trade-off as a function of $\beta$, where the total CBR $\rho = (R + M/N) / m$.}
	
	\begin{threeparttable}
		\begin{tabular}{m{0.6cm}|m{1.1cm}|m{1.6cm}|m{1.6cm}|m{1.7cm}}
			
			\Xhline{1pt}
			
			\centering 	\multirow{2}*{$\beta$} & \centering \multirow{2}*{\shortstack{PSNR\\(dB)}} & \centering \multirow{2}*{\shortstack{$\rho$\\(total CBR)}} & \centering \multirow{2}*{\shortstack{$R/m$\\(data CBR)}}  & \centering \multirow{2}*{\shortstack{$M/(mN)$\\(model CBR)}} \tabularnewline
			
			~ & ~ & ~ & ~  & ~ \tabularnewline
			
			\hline\hline
			
			\centering -- & \centering 31.80 & \centering 0.04967 & \centering 0.04967 & \centering 0   \tabularnewline
			
			
			\centering 16 & \centering 33.13 & \centering 0.04870 & \centering 0.04860 & \centering 0.00011   \tabularnewline
			
			
			\centering 4 & \centering 33.33 & \centering 0.04853 & \centering 0.04825 & \centering 0.00027  \tabularnewline
			
			
			\centering 1 & \centering 33.65 & \centering 0.04885 & \centering 0.04811 & \centering 0.00073  \tabularnewline
			
			
			\centering 0.1 & \centering 34.35 & \centering 0.05245 & \centering 0.04867 & \centering 0.00378  \tabularnewline
			
			
			\centering 0.01 & \centering 35.21 & \centering 0.07377 & \centering 0.04549  & \centering 0.02828  \tabularnewline
			
			\Xhline{1pt}
			
		\end{tabular}
	\end{threeparttable}
	\label{table_R-D-M_tradeoff}
\end{table}

\begin{figure}[t]
	\centering
	{
		\includegraphics[scale=0.5]{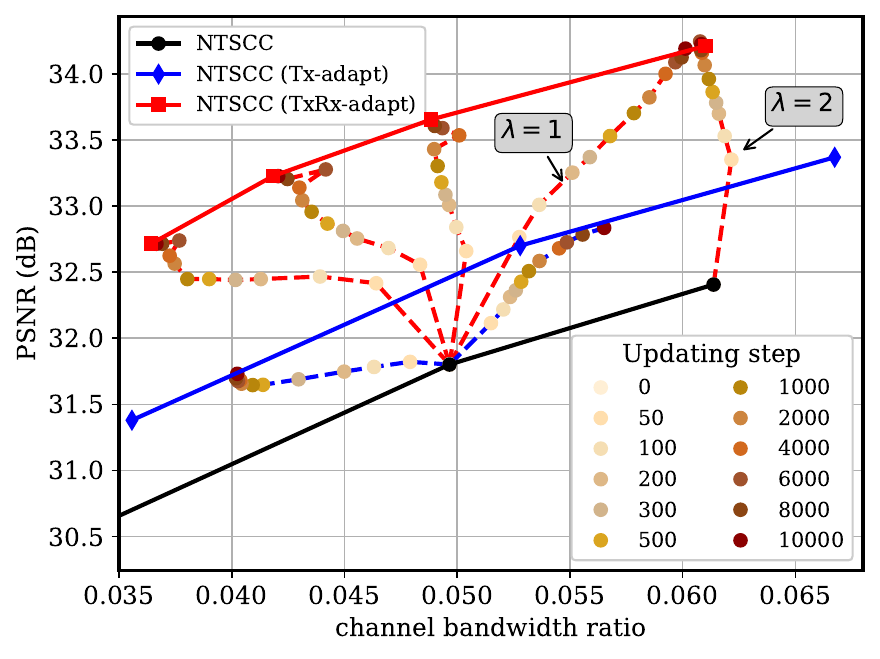}
	}
	\caption{Online learning on the \texttt{game} video under AWGN channel at $\text{SNR} = 10$dB. Each dot shows performance under different updating steps.}
	\label{Fig_online_learning}
\end{figure}

We extend our analysis by examining the impact of $\lambda$ in domain adaptation. Fig. \ref{Fig_online_learning} depicts the model updating progress on the CBR-PSNR plot with various $\lambda$ values and a fixed $\beta=1$. From the results, we observe that the significant gains from domain adaptation are achieved during the early stages of model updating, within just hundreds of iterations, and the performance continues to improve moderately during online updating. Additionally, optimizing for different $\lambda$ values can cause the R-D-M curve to move in different directions, but the final R-D-M performance point seems independent of the starting point from baseline model, as seen in the highlighted $\lambda=1$ and $\lambda=2$ curves. This phenomenon suggests that, given the model stream weight $\beta$, the maximum reduction in the amortization gap on the receiver has almost been reached. To further reduce the amortization gap, the only approach is to limit the range of the adaptation domain.

\subsubsection{Performance-Complexity Tradeoff}

We present the average end-to-end processing latency and the coding efficiency of considered coded transmission schemes on the 480p VSD4K dataset in Table \ref{table_latency}. This includes coding latency and the BD-$\rho$ (``$\downarrow$'' means smaller is better), which measures bandwidth ratio savings at the equivalent quality, ``$+/-$'' signs indicate more/fewer bandwidth cost than baseline \cite{bjontegaard2001calculation}. All experiments are conducted using PyTorch 1.9.0, with the Inter Xeon Gold 6226R CPU (mainly for arithmetic codec) and one RTX 3090 GPU (for model updating and inference, and LDPC codec).

\begin{table}[t]
	\renewcommand{\arraystretch}{1.3}
	\centering
	\small
	
	\caption{Transmission efficiency and averaged encoding/decoding latency comparison. }
	
	\begin{threeparttable}
		\begin{tabular}{m{2.9cm}|m{0.75cm}|m{0.75cm}|m{1.4cm}|m{0.8cm}}
			
			\Xhline{1pt}
			
			\multirow{2}*{\shortstack{Transmission scheme}} & \multicolumn{2}{c|}{\shortstack{BD-$\rho$ (\%, $\downarrow$)}} & \multicolumn{2}{c}{End-to-end latency} \tabularnewline
			\cline{2-5}
			~ & \centering 0dB & \centering 10dB &  \centering Enc. & \centering Dec. \tabularnewline
			\hline
			\hline
			
			BPG + 5G LDPC & \centering 13.1 & \centering 1.1 & \centering 3.8s & \centering 380ms \tabularnewline
			
			VTM + 5G LDPC & \centering --1.2 & \centering --17.7 & \centering 114s & \centering 430ms \tabularnewline
			
			NTC \cite{balle2018} + 5G LDPC & \centering 22.9 & \centering 11.2 & \centering 27ms & \centering 60ms \tabularnewline	
			
			\hline
			
			Deep JSCC \cite{DJSCCF} & \centering 42.3 & \centering 46.4 & \centering 17ms & \centering 25ms \tabularnewline	
			
			NTSCC \cite{dai2022nonlinear} & \centering 0 & \centering 0 & \centering $t_0=25$ms & \centering 15ms \tabularnewline
			
			\hline
			
			Tx-adapt, instance & \centering --9.0 & \centering --9.0 & \centering $t_1$  & \centering 15ms \tabularnewline
			
			TxRx-adapt, domain & \centering --19.2 & \centering --16.4 & \centering $t_2$ or $t_0$  & \centering 15ms \tabularnewline
			
			\Xhline{1pt}
		\end{tabular}
		\begin{tablenotes}
			\footnotesize
			\item[1] The Tx-adapt encoding time $t_1$ is associated with the total number of instance adaptation steps. As a reference, it takes about 29.7s for Tx model updating in Algorithm \ref{alg_tx_adp_model} with $T_{\max}=100$ steps and 18.3s for updating latents in Algorithm \ref{alg_tx_adp_code} with $Y_{\max}=S_{\max}=50$ steps.
			\item[2] The TxRx-adapt online learned NTSCC system requires online updating full-model parameters before inference. As a reference, it takes about $t_2=2.2$h for $T_{\max}=10000$ steps. After updating, the subsequent encoding latency is same as the standard NTSCC, i.e., $t_0=25$ms.
		\end{tablenotes}
	\end{threeparttable}
	\label{table_latency}
\end{table}

Results show that learned end-to-end data transmission schemes (e.g., Deep JSCC and NTSCC) run much faster than classical layered designed schemes. This makes emerging end-to-end data transmission schemes particularly suitable for low-latency applications such as live video delivery, which requires real-time encoding on the client side. We then discuss the online learning schemes. Essentially, they enhance the coding efficiency of end-to-end data transmission schemes by leveraging computation at the transmitter. For the instance adapt scheme, it requires tens of seconds of additional encoding time per frame for updating codes or encoder model, in exchange for 9\% bandwidth saving compared to the standard NTSCC. The instance adapt paradigm can potentially support wireless image/video delivery services, where encoding can be run in parallel and in advance.

For the TxRx domain adaptation scheme, in order to achieve improved coding efficiency, it is necessary to perform online updating of the full-model parameters and transmit the model stream to the receiver before initiating the encoding process. Consequently, there will be a delay before the full-model updating is completed. If the updating process is unacceptable for delay-sensitive applications (e.g., live video delivery), we recommend reverting to the baseline model for a while before the updating process is completed. In this way, one can ensure low end-to-end latency and achieve additional gains from domain adaptation in the same time.

\begin{figure}[t]
	\centering
	{
		\includegraphics[scale=0.5]{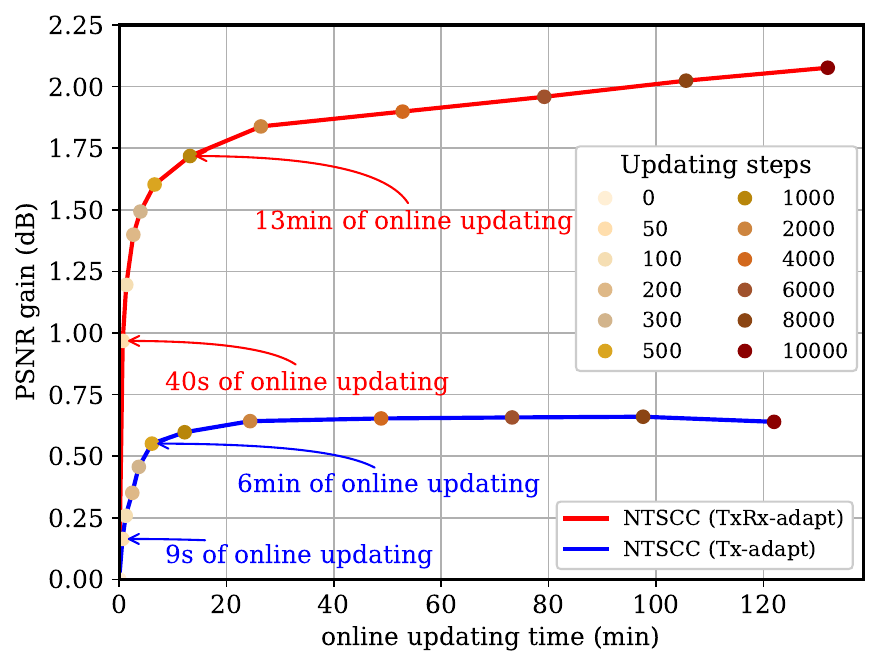}
	}
	\caption{The PSNR gain versus standard NTSCC over time during model updating. Each dot shows performance under different updating steps.}
	\label{Fig_psnr_gain_vs_time}
\end{figure}

Another advantage of the domain adaptation paradigm is that online fine-tuned model can converge more rapidly if the test-time content is relatively consistent, such as the scenarios including a fixed-location camera, cityscapes, live broadcasts, or video conferences. In Fig. \ref{Fig_psnr_gain_vs_time}, we plot the detailed PSNR gain versus the baseline over the online training time on the \texttt{game} sequence. As seen in this figure, about 50\% of the gain can be achieved after updating for less than 1 minute, and 80\% of the gain can be attained with about 10 minutes of updating. Since this is a ``one-time'' cost amortized across all the video frames, the additional training cost is negligible compared to the channel bandwidth saving benefits it brings, especially for long videos.

\subsubsection{Results for Unseen Data}

\begin{figure}[t]
	\centering
	{
		\includegraphics[scale=0.5]{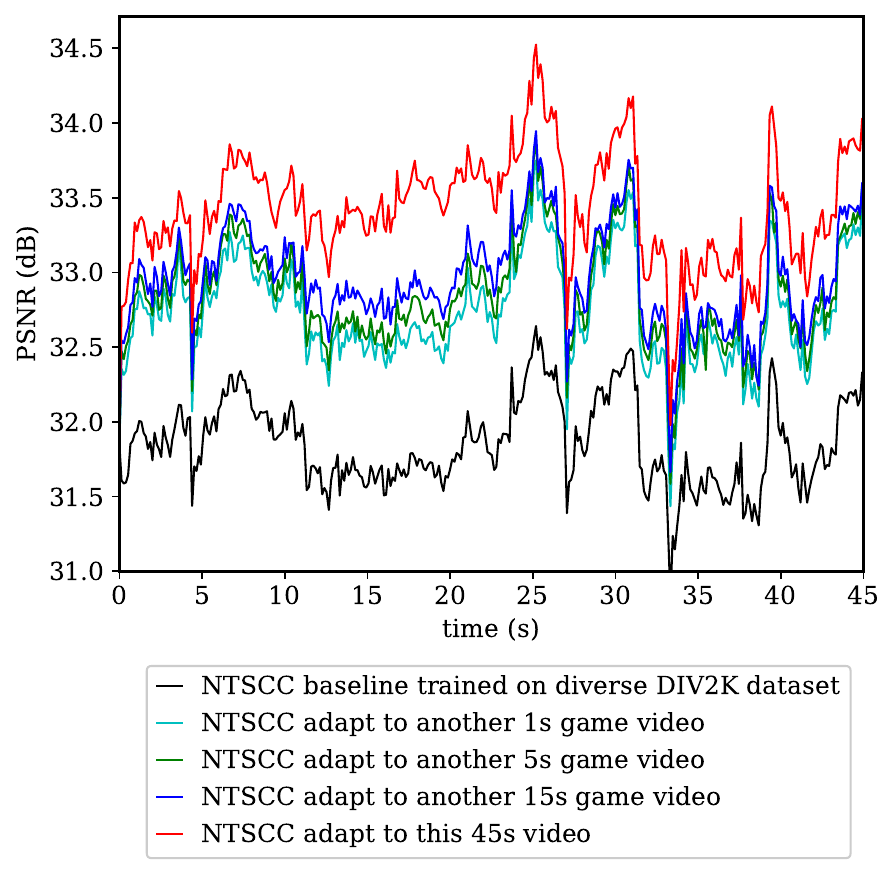}
	}
	\caption{Case study: live video delivery. We downloaded a new 45s video of the same game from YouTube and used it to simulate a live video delivery scenario, where the video content to be encoded cannot be known in advance. The bandwidth costs of different schemes  are adjusted to be the same.}
	\label{Fig_noisy_prior}
\end{figure}

To evaluate the performance of our full-model adaptation method on unseen data, we downloaded a new 45s video from YouTube as test sequence, which contains the same game as the \texttt{game} sequence in the VSD4K dataset. We use the new video as unseen data to simulate a live video delivery scenario, where the video content to be encoded cannot be known in advance. In other words, the online learning can only utilize previously collected video content. The results in Fig. \ref{Fig_noisy_prior} demonstrate that the proposed online learning method can yield significant performance gains compared to the baseline, even when the training set consists of a short video sequence with a duration of only 1s. Moreover, the models adapt to another game video can still approximate the performance of the best models adapted to this precise video. This observation indicates that the proposed method exhibits good robustness to previously unseen data from the same category, as the model assimilates domain-specific knowledge through the online learning process.

\subsubsection{Result for Various Video Resolution}

\begin{figure}[t]
	\centering
	{
		\includegraphics[scale=0.5]{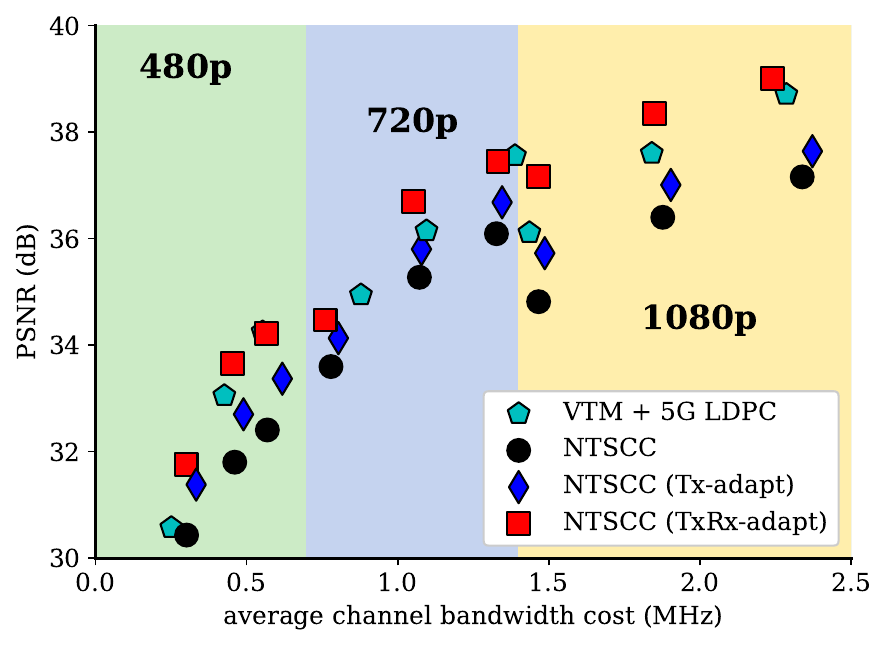}
	}
	\caption{PSNR performance of transmitting the \texttt{game} video at difference resolutions over the AWGN channel at $\text{SNR}=10$dB, where the average bandwidth cost is computed within 100ms according to the 5G OFDM configurations: 14 OFDM symbols@1ms with the subcarrier bandwidth 15KHz.}
	\label{Fig_diverse_resolution}
\end{figure}

We further report the performance on the \texttt{game} video of different resolutions in Fig. \ref{Fig_diverse_resolution}. These results validate the stability of quality improvements achieved using our proposed overfitting mechanism. The proposed online learned NTSCC shows superior performance over the SOTA VTM + 5G LDPC scheme, particularly for high-resolution video I-frames.

\subsection{Results Discussion}

Regarding the above experimental results and ablation study, we can draw two conclusions:
\begin{itemize}
  \item First, our online overfitting mechanism enables semantic communication system to efficiently adapt to both source content and wireless channel state. It can greatly reduce the model amortization gap, which finally contributes to the system performance gain.

  \item Second, the small additional complexity caused by model adaptation involves only at the transmitter. There is no extra complexity at the receiver, which is friendly to most content delivery tasks with the request of low decoding latency. It is aligned with the evolution idea of both traditional and neural video codec.
\end{itemize}

In summary, as mentioned in Section \ref{section_introduction}, almost all traditional source compressors follow the hybrid transform coding paradigm to evolve, e.g., mode selection in HEVC \cite{sullivan2012overview} and VVC \cite{bross2021developments}. Accordingly, the idea of signal-dependent transform in traditional source compression methods inspires us to upgrade the deep learning based semantic communication system to a content-channel-dependent adaptive mode. The marriage of deep learning method and traditional coding idea can bring significant performance improvement.

Based on previous works about semantic communications \cite{DJSCC,DJSCCF,DJSCCL,jankowski2020wireless,JSCCtext,choi2019neural,xu2021wireless,wang2021novel,dai2022nonlinear,yang2022deep}, and the work in this paper, we think that the advantages of neural network based end-to-end semantic communication system are three folds:
\begin{itemize}
  \item First, the excellent content and semantics adaptivity of neural network is superior to signal processing based traditional models since the network parameters are learned based on lots of practical source and channel data samples while the models in the SOTA coded transmission standards are handcrafted based on prior knowledge.

  \item Second, the neural network can well represent and utilize source and channel features, which makes the semantic communication system can be optimized toward both human view perception and machine vision tasks. However, the existing source and channel coding standards only pursue high performance toward the objective quality assessment indices, e.g., PSNR.

  \item Third, the R-D optimization guided neural network training and adaptive switching for semantic communication is quite effective and efficient. As analyzed in this paper, a single model to deal with all the source data with diverse structures and varying wireless channel states is inefficient obviously. Therefore, the adaptively learning and switching according to source data and channel state is \emph{a necessary solution} to enhance the R-D performance for all deep learning based semantic communication systems. In addition, compared with the adaptive coding paradigm used in traditional source and channel coding where one appropriate coding mode is selected from the predefined mode pool with limited number of options, this paper leverages the overfitting property of neural network to adapt to arbitrary coding mode. Hence, the proposed adaptive semantic communication system is much more flexible to be tailored for specific source and channel instance with lower complexity.
\end{itemize}

In a nutshell, our overfitting method has efficiently catalyzed semantic communication system to provide more promising results. It is a necessary approach for all learning-based end-to-end communication system to further boost performance, which is indeed aligned with the evolution route of both traditional and neural video codec \cite{ma2019image}.

\section{Conclusion}\label{section_conclusion}

In this paper, we have proposed an online learned end-to-end data transmission paradigm by well leveraging the deep learning model's overfitting property. Our adaptive NTSCC system can for instance or domain be updated after deployment, which leads to substantial gains on the bandwidth ratio-distortion performance. In this way, the emerging semantic communication is further upgraded to the adaptive semantic communication system. Specifically, we have proposed several lightweight methods to update the learned data transmission models. The ingredients of wireless transmitted stream include both the semantic representations of source data and the updated decoder model parameters. Accordingly, we have formulated the new optimization problem whose goal is minimizing the loss function that is a tripartite trade-off among data stream rate, model stream rate, and end-to-end distortion terms. Results have verified the substantial gains of the online learned NTSCC system, whose performance has surpassed the SOTA engineered coded transmission systems. As a new paradigm, our model adaptation methods have the potential to catalyze semantic communication upgrading to a new era.

\bibliographystyle{IEEEbib}
\bibliography{myRef}

\end{document}